\documentclass{aa}
\usepackage{graphicx}
\usepackage{amsmath, amssymb}
\usepackage{bm}
\usepackage[varg]{txfonts}
\usepackage{textcomp}
\usepackage{hvfloat}
\usepackage{natbib}
\usepackage{hyperref}

\bibpunct{(}{)}{;}{a}{}{,}

\begin{document}

\newcommand{\micron}{\textmu m}		

\title{Vertical settling and radial segregation of large dust grains in the circumstellar disk of the Butterfly Star}

\author{C.~Gr\"afe\thanks{e-mail: cgraefe@astrophysik.uni-kiel.de} \inst{\ref{inst1}}
	\and S.~Wolf \inst{\ref{inst1}}
	\and S.~Guilloteau \inst{\ref{inst2}, \ref{inst3}}
	\and A.~Dutrey \inst{\ref{inst2}, \ref{inst3}}
	\and K.~R.~Stapelfeldt \inst{\ref{inst4}}
	\and K.~M.~Pontoppidan \inst{\ref{inst5}}
	\and J.~Sauter \inst{\ref{inst1}}
       }

\institute{
	    University of Kiel, Institute for Theoretical Physics and Astrophysics, Leibnizstrasse 15, 24118 Kiel, Germany 	\label{inst1}
	    \and Univ. Bordeaux, LAB, UMR 5804, F-33270, Floirac, France							\label{inst2}
	    \and CNRS, LAB, UMR 5804, F-33270 Floirac, France									\label{inst3}
	    \and Exoplanets and Stellar Astrophysics Laboratory, NASA Goddard Space Flight Center, Greenbelt, MD 20771, USA 	\label{inst4}
	    \and Space Telescope Science Institute, Baltimore, MD 21218, USA 							\label{inst5}
	  }

\date{Received 12 November 2012 / Accepted 13 March 2013}

\abstract
  {Circumstellar disks are considered to be the environment for the formation of planets. The growth of dust grains in these disks is the first step in the core accretion-gas capture planet formation scenario. Indicators and evidence of disk evolution can be traced in spatially resolved images and the spectral energy distribution (SED) of these objects.}
  {We develop a model for the dust phase of the edge-on oriented circumstellar disk of the \object{Butterfly Star} which allows one to fit observed multi-wavelength images and the SED simultaneously.}
  {Our model is based on spatially resolved high angular resolution observations at 1.3\,mm, 894\,\micron, 2.07\,\micron, 1.87\,\micron, 1.60\,\micron, and 1.13\,\micron\ and an extensively covered SED ranging from 12\,\micron\ to 2.7\,mm, including a detailed spectrum obtained with the Spitzer Space Telescope in the range from 12\,\micron\ to 38\,\micron. A parameter study based on a grid search method involving the detailed analysis of every parameter was performed to constrain the disk parameters and find the best-fit model for the independent observations. The individual observations were modeled simultaneously, using our continuum radiative transfer code.}
  {We derived a model that is capable of reproducing all of the observations of the disk at the same time. We find quantitative evidence for grain growth up to $\sim\!100\,$\micron-sized particles, vertical settling of larger dust grains toward the disk midplane, and radial segregation of the latter toward the central star. Within our best-fit model the large grains have a distribution with a scale height of $3.7\,$AU at $100\,$AU and a radial extent of $175\,$AU compared to a hydrostatic scale height of $6.9\,$AU at $100\,$AU and an outer disk radius of $300\,$AU. Our results are consistent with current theoretical models for the evolution of circumstellar disks and the early stages of planet formation.}
  {}


\authorrunning{C.~Gr\"afe et al.}

\maketitle

\section{Introduction}
\label{section:introduction}

Circumstellar disks are a natural outcome of the star formation process and are known to dissipate over time \citep[$\sim\!\!10^6$-$10^7\,$yr;][]{Haisch01} by means of stellar winds or by photoevaporation caused by either the radiation of the central star or by an external source of radiation \citep[e.g.,][]{Hollenbach00, Clarke01, Alexander07}, by accretion onto the central star \citep{Hartmann98}, by grain growth and fragmentation \citep[e.g.,][]{Dullemond05, Dominik07, Natta07, Birnstiel11, Sauter11, Garaud12, Ubach12}, and by the formation of planets \citep[e.g.,][]{Pollack96, Boss02, Armitage07, Armitage10, Fortier12}. The way in which the disk material dissipates has important implications for the possibility of planet formation.
\par
During the last decades several theories to explain the formation of planets in circumstellar disks have been proposed, for example, the core accretion-gas capture scenario \citep{Pollack96, Papaloizou06, Lissauer07}. Sub-micron-sized dust particles, which are strongly coupled to the motion of the gas in the disk via gas drag, agglomerate through low-velocity collisions to eventually form planetesimals of several kilometers in size \citep[e.g.,][]{Beckwith00, Dominik07, Natta07}. During this coagulation process, larger particles decouple from the turbulent gas motion, settle toward the disk midplane and radially drift toward the inner parts of the disk as a result of the impact of stellar gravity and gas drag \citep[e.g.,][]{Weidenschilling77, BarriereFouchet05, Fromang06, DAlessio06}. The growth beyond planetesimals occurs via direct collisions, with an increasing role for gravitational focusing as masses become larger, leading to the gravitational agglomeration of these bodies to rocky planets \citep{Safronov69}. A phase of runaway growth occurs, followed by an oligarchic growth phase \citep{Kokubo98, Thommes03}.
\par
There are, however, many unresolved problems in the picture of this formation process. For example, it is uncertain how dust grains overcome the radial drift barrier \citep{Weidenschilling77}. This occurs as it is expected that meter-sized bodies migrate toward the star in a timescale that is shorter than the timescale for further growth which effectively stops further grain growth. Another example is the fragmentation barrier, where the boulders are destroyed at typical collision speeds, halting the dust growth at centimeter to meter sizes \citep{Benz00, Blum08, Brauer08}. Moreover, the dust coagulation process, the first stage of planet formation, must also overcome the charging barrier \citep{Okuzumi08} and bouncing barrier \citep{Zsom10} which affect the later stages, too.
\par
Besides the hypothesis that planetesimals may form from pairwise collisional growth of smaller bodies alone, there is the hypothesis that, when the dust particles have reached cm-scale as a consequence of this collisional growth, the planetesimals may form from the gravitational fragmentation of a dense particle sub-disk near the equatorial plane resulting from the dust settling. Protoplanetary disks may support regions within which turbulence acts to locally enhance the ratio of solids to gas. As a result, patches of the disk may become dense enough to trigger a gravitational instability and collapse into planetesimals \citep{Safronov69, Goldreich73, Armitage07, Chiang10}. This suggestion entirely bypasses the size scales that are most vulnerable to radial drift and would allow the radial drift and fragmentation barrier to be overcome.
\par
With the current observational capacities it is not possible to directly observe large dust boulders ($>\!1\,$m) and planetesimals, but signs and implications of their formation can be detected. The stratified structure resulting from grain growth and settling has a significant impact on observable quantities of the disk \citep[e.g.,][]{Dullemond04}. The spectral energy distribution (SED) provides constraints on the dust mass and grain-size distribution. Scattered light images at near-infrared (NIR) and mid-infrared (MIR) wavelengths probe the surface layers of the optically thick disks and reveal properties of small dust grains through wavelength-dependent opacity \citep[e.g.,][]{Watson07}. Continuum observations at submillimeter and millimeter (mm) wavelengths are mainly sensitive to the bulk of the disk mass closer to the midplane and play an important role in constraining disk structure properties, for example, the spatial dust distribution of larger grains \citep{Andrews05, Andrews07}. The analysis of scattered light or re-emission images or the SED only provides a limited view of a disk leading to ambiguities in the derived disk and dust properties \citep[e.g.,][]{Chiang01}. To precisely study the physical processes like dust evolution and to strongly reduce model degeneracies, it is essential to combine resolved images at different wavelengths and a SED with a good wavelength coverage in a multi-wavelength observational and modeling approach.
\par
In this paper we study the circumstellar disk of the prominent Butterfly Star, also known as IRAS 04302+2247, using this approach. This Class~I young stellar object \citep[YSO; e.g.,][]{AdamsLadaShu87, Lada87} is located in the Taurus-Auriga molecular cloud complex at a distance of $\sim\!140$\,pc \citep{Kenyon94}. It is surrounded by an edge-on seen flared circumstellar disk, so the central star is not visible directly \citep[inclination $i = 90\degr\pm3\degr$;][]{Padgett99, Padgett01, Wolf03}. As shown by interferometric observations at submm and mm wavelengths, the disk exhibits a radius of about 300\,AU \citep{Wolf03, Wolf08}. Several attempts have been undertaken to model the structure and physical conditions in the circumstellar environment of this object \citep[e.g.,][]{Lucas97, Wolf03, Stark06}. In particular, \citet{Wolf03} were the first to base their modeling on scattered light images and resolved mm maps and, therefore, were the first to consider wavelength regimes tracing different physical processes in different regions of the circumstellar environment. They concluded that the grains in the outer parts of the circumstellar environment are comparable to those from the interstellar medium (ISM), while dust grains have grown via coagulation by several orders of magnitude in the much denser parts of the disk. However, the observational data they presented did not allow them to constrain the spatial dependence of the dust grain properties in the disk interior. 
\par
We present new observations and a coherent multi-wavelength model for the circumstellar disk of the Butterfly Star that accounts for spatially resolved data sets ranging from the NIR to mm wavelengths as well as for the well-sampled SED of this object. These observations provide different, complementary detailed perspectives on the disk structure and dust properties. Here, coherent means that the model is capable of reproducing all considered observations obtained at different wavelengths at the same time. This new modeling became necessary as new observations are not properly reproduced by the model of \citet{Wolf03, Wolf08}. Furthermore, these new observations with much higher angular resolution promise to allow one to put stronger constraints on the disk structure and dust grain properties.
\par
The observations and data reduction are introduced in Sect.\,\ref{section:obs_datared}. In Sect.\,\ref{section:basisformodeling}, we characterize the data set that is the basis for our modeling. A detailed description of our model can be found in Sect.\,\ref{section:modeldescription} and the entire modeling process is described in Sect.\,\ref{section:modeling}. The results of this study are presented in Sects.\,\ref{section:firstmodelingstep} and \ref{section:secondmodelingstep} and their implications are discussed in Sect.\,\ref{section:discussion}. Finally, Sect.\,\ref{section:summary_conclusion} contains a summary and concluding remarks.
\par

\section{Observations and data reduction}
\label{section:obs_datared}

The data set used in this study includes both spatially resolved images at mm, submm, and NIR wavelengths, as well as a well-sampled SED. In this section these observations and the data reduction procedures are discussed.
\par

  \subsection{Millimeter observation}
  \label{subsection:mm_obs}

The Butterfly Star was observed with the IRAM Plateau de Bure interferometer \citep[PdBI,][]{PdBI91} in its two most extended configurations (A and B). The dual polarization receivers were tuned near 220 GHz ($\lambda = 1364\,$\micron). The wideband WIDEX correlator provided a total bandwidth of 4 GHz in each of the two polarizations. The covered frequency range included lines of the $^{13}$CO and C$^{18}$O isotopologues, as well as H$_2$CO and SO. Spectral line results will be reported in a separate paper, but the continuum data discussed here was generated by avoiding line contamination.
\par
Observations were performed on February 3 and 4, 2010 (Aq configuration, two baselines of 730-760\,m, excellent RMS (baseline-based) phase noise $<35\degr$), January 22 (Aq configuration), and February 10, 2011 (B configuration,  up to 450\,m baselines, RMS phase noise $<45\degr$). Although the repeated configurations produced similar results, data from January 22, 2011 was not included in the final analysis because of its higher phase noise. With very good observing conditions ($< 1\,$mm water vapor, going up to 1.5\,mm only for the last 2 hours of February 10, 2011), single sideband system temperatures ranged between 100 to 140\,K. Phase noise yields an effective seeing of approximately 0.15$\arcsec$.
\par
Phase calibration was performed using the nearby quasars 0400+258 and 0507+179, and flux calibration is referred to MWC\,349, with an assumed flux density of 1.7\,Jy at this frequency, with an overall calibration accuracy of order 10\,\%.
\par
With natural weighting, the angular resolution of the data is $0.59\arcsec\times0.35\arcsec$ at a position angle ($PA$) of $31.1\degr$. The shortest baseline is large, 79\,m including projection effects, so that structures larger than about 4$\arcsec$ could be substantially resolved out. However, given the source elongation and orientation and the beam shape, deconvolution is efficient in recovering most structures.
\par
The total flux density is about $101\,{\rm mJy}\pm10\,$mJy. The expected thermal noise in the map is 0.18\,mJy/beam, but because of phase noise, the dynamic range is limited and the effective noise level in the deconvolved map is $\sim\!0.4\,$ mJy/beam. 
\par

  \subsection{Submillimeter observation}
  \label{subsection:submm_obs}

Observations of the Butterfly Star with the Submillimeter Array \citep[SMA,][]{SMA04} were carried out on January 9, 2006 using the upper and lower sidebands at 330\,GHz and 340\,GHz, respectively ($\lambda = 894\,$\micron). The angular resolution of the data is $0.67\arcsec\times0.53\arcsec$ at $PA = -76.5\degr$. The SMA observations and the data reduction were described in detail by \citet{Wolf08}.
\par

  \subsection{Near-infrared observations}
  \label{subsection:nir_obs}

The NIR observations of the Butterfly Star were obtained with the Near Infrared Camera and Multi-Object Spectrometer \citep[NICMOS,][]{NICMOS98} on the Hubble Space Telescope (HST) on August 19, 1997. The NICMOS NIR Camera, NIC2, and the filters F110W ($\lambda_{\rm c} = 1.13\,$\micron), F160W ($\lambda_{\rm c} = 1.60\,$\micron), F187W ($\lambda_{\rm c} = 1.87\,$\micron), and F205W ($\lambda_{\rm c} = 2.07\,$\micron) were used. The data reduction and flux calibration were described in detail by \citet{Padgett99}. Figure~\ref{Fig:BS_RGB} shows a composite image of the Butterfly Star at NIR wavelengths.
\par

  \subsection{Spectroscopic data}
  \label{subsection:spec}

The Butterfly Star was observed with the low-resolution modules of the Spitzer Space Telescopes (SST) InfraRed Spectrograph (IRS) on October 1, 2007 and March 1, 2004 for short-low (SL) and long-low (LL), respectively. The LL observation was obtained as part of the IRS guaranteed time observations (PI: Houck), while the deeper SL observations were obtained as part of the open time program 30765 (PI: Stapelfeldt). The total exposure time  was 587 seconds for the SL modules and 12.6 seconds for the LL modules. The two-dimensional droop-corrected frames obtained using pipeline version S16.1.0 were co-added for each nod position, and the two nods were pairwise differenced to remove the background. One-dimensional spectra were extracted using apertures of 3 pixels for the SL modules, while 4 and 5 pixel apertures were used for the LL2 and LL1 modules, respectively. The one-dimensional spectra were flux-calibrated using the spectral response function included in the SST pipeline.
\par

\section{Basis for modeling}
\label{section:basisformodeling}

The basis of our modeling forms the (sub)mm and NIR maps, as well as the SED, including the SST/IRS spectrum. This data set is briefly characterized in this section.
\par

  \subsection{Images}
  \label{subsection:basis_img}

\begin{figure}[h]
 \centering
  \includegraphics[width=0.49\textwidth]{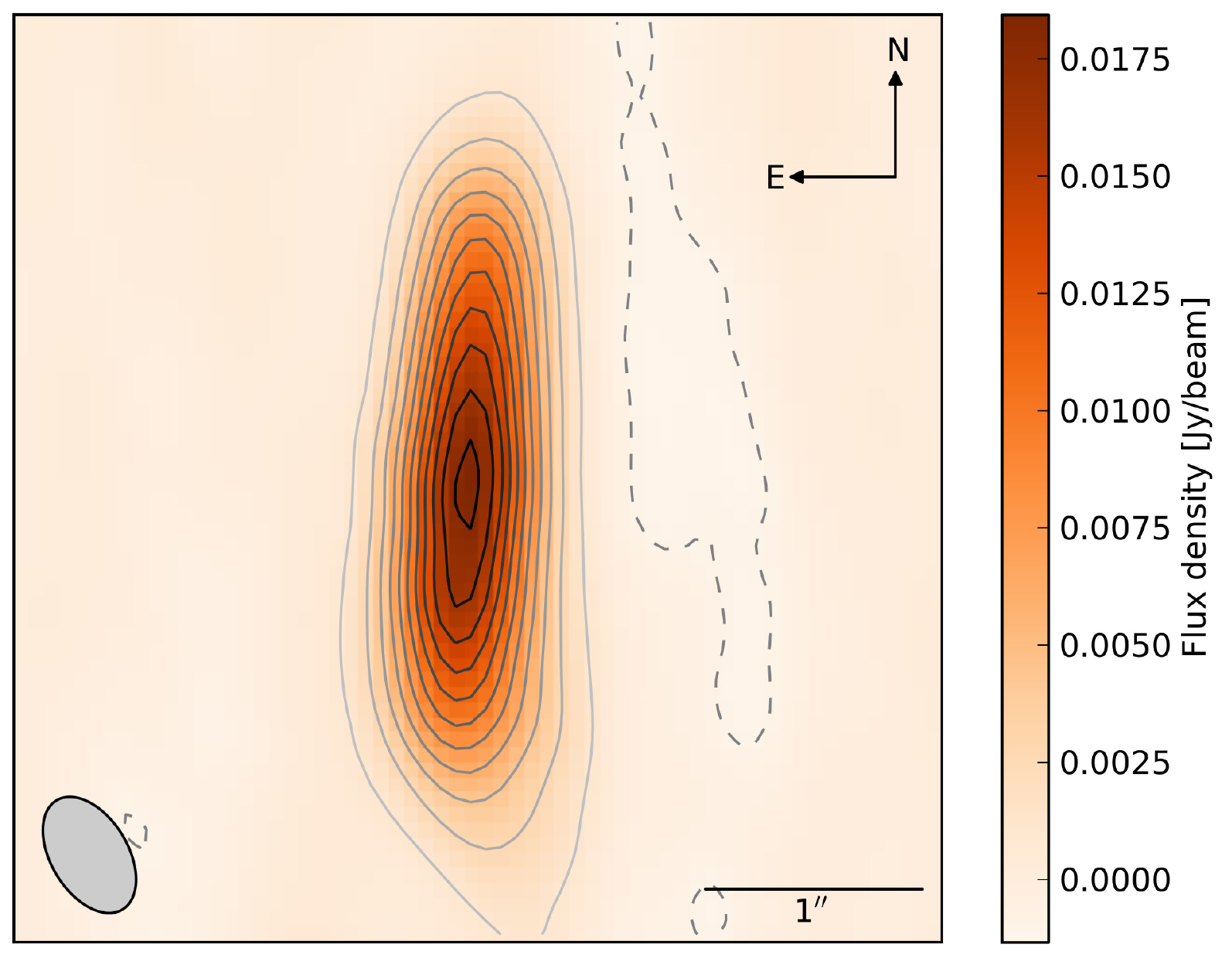}
\caption{$1.3$\,mm map obtained with the PdBI. Negative contours are dashed and at -3\,$\sigma$, where $\sigma$ is the background noise of the map. The positive contour levels are in steps of 5\,$\sigma$ starting at 3\,$\sigma$ and going up to 53\,$\sigma$. The ellipse in the bottom left illustrates the shape and orientation of the beam.}
\label{Fig:PdBI_1}
\end{figure}

\begin{figure}[h]
 \centering
  \includegraphics[width=0.49\textwidth]{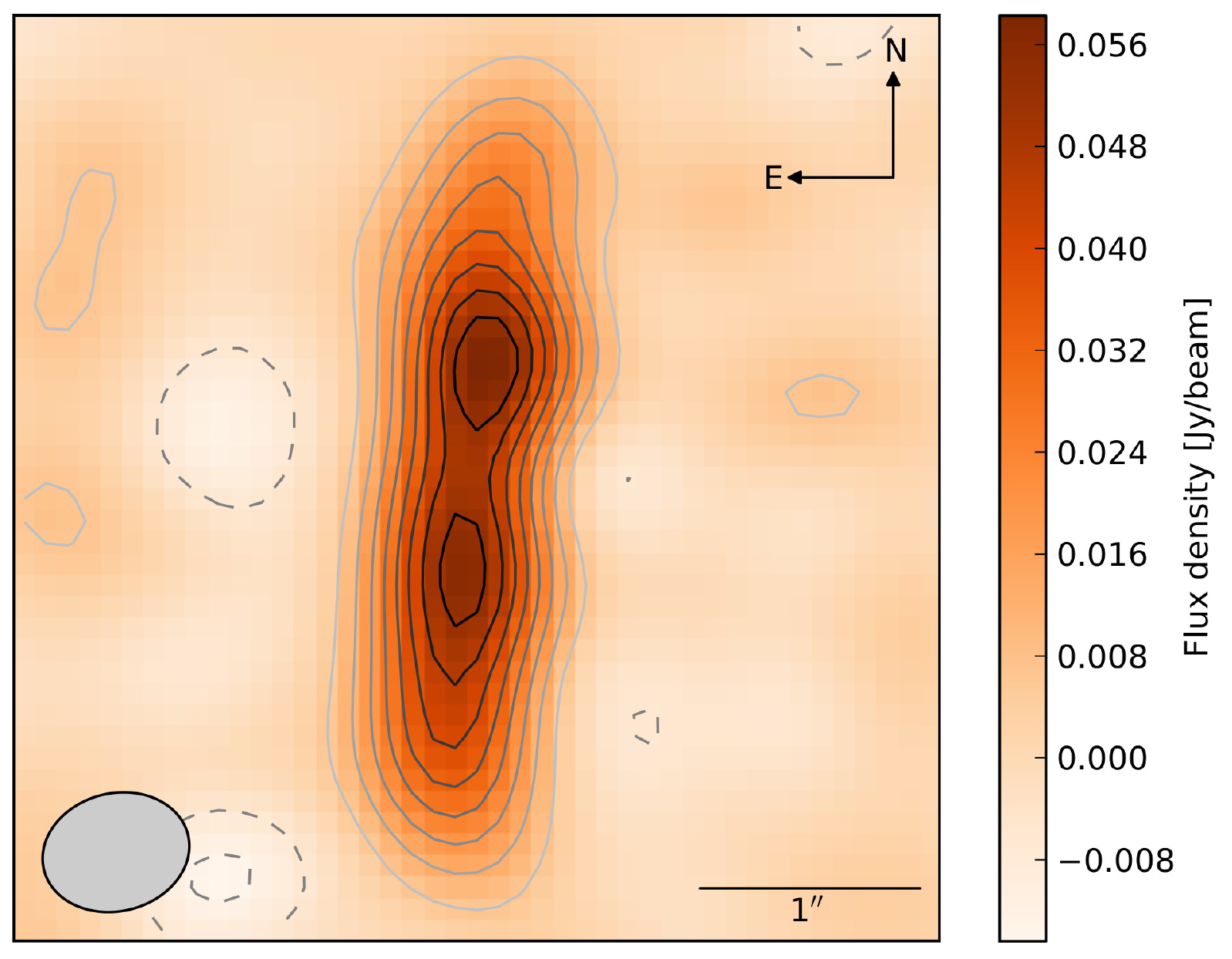}
\caption{$894$\,\micron\ map obtained with the SMA. The contour levels are in steps of 2\,$\sigma$ starting at -4\,$\sigma$, leaving out the zero contour level, and going up to 16\,$\sigma$. The ellipse in the bottom left illustrates the shape and orientation of the beam.}
\label{Fig:SMA_1}
\end{figure}

The interferometric continuum maps at $1.3$\,mm and $894$\,\micron\ are shown in Figs.~\ref{Fig:PdBI_1} and \ref{Fig:SMA_1}. The emission seen in these maps is thermal re-emission of the dust. While the source is unresolved vertically to the disk in both maps, it is well-resolved in radial direction. The circumstellar disk seen in both maps has a north/south extension of $\sim\!600$\,AU. While the brightness structure of the $1.3$\,mm map shows a clear maximum at the center, a local minimum is visible at this position in the $894$\,\micron\ map. The nature of this local minimum is discussed in Sect.\,\ref{subsection:minimum_submm}.
\par
\begin{figure}[h]
 \centering
  \includegraphics[width=0.49\textwidth]{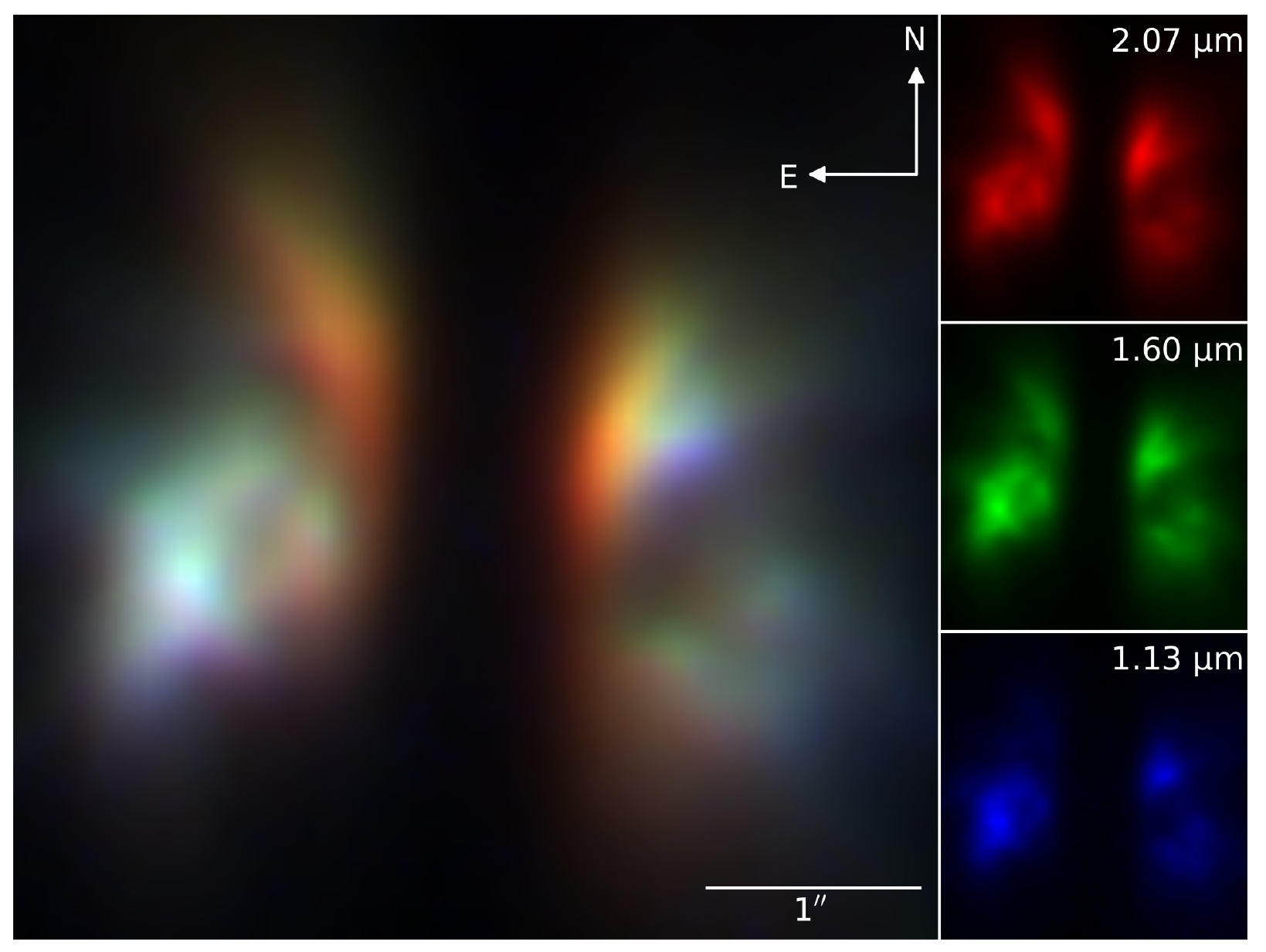}
\caption{Composite image of the Butterfly Star seen in scattered light. For details see Sects.\,\ref{subsection:basis_img} and \ref{subsection:secondmodelingstep_results}.}
\label{Fig:BS_RGB}
\end{figure}

The emission seen in the NIR images is scattered light from the central star (see Fig.~\ref{Fig:BS_RGB} below and Fig.~1 in \citet{Wolf03}). Their appearance is dominated by a totally opaque linear feature that bisects the scattered light structure. This feature is interpreted as a circumstellar disk seen edge-on. Because of the large optical depth of this disk, no point source is detected at any of the observed NIR wavelengths. The dependence of the height of the disk on the wavelength is clearly visible. The bipolar scattered light nebula, which extends $\sim\!900$\,AU in the direction north/south, also shows a complex morphology with approximately equal brightness between the eastern and western lobes. As this appearance reminds one of a butterfly, the name {\it Butterfly Star} has been established \citep{Lucas97}. Comparing the (sub)mm maps with the NIR images it can be seen that the position and orientation of the elongated (sub)mm structure perfectly fits that of the dark lane in the NIR.
\par
We rotate the observed images by the position angle of the image $y$-axis and the position angle of the Butterfly Star ($PA=175\,\degr$, as simulated images are symmetric we use $PA=-5\,\degr$) in order to align the major axis of the dust lane with the vertical axis, as it is for the simulated images.

  \subsection{Spectral energy distribution}
  \label{subsection:basis_sed}

The photometric measurements used in this modeling are presented in Table~\ref{Tab:fluxes}. In addition, the SST/IRS spectrum in the range from 12\,\micron\ to 38\,\micron\ is considered. In our disk modeling we do not include the scattered-light part of the SED below 12\,\micron\ as it is not an objective of this study to reproduce the highly-structured envelope seen in the NIR images (see Sect.\,\ref{subsection:modelfittingproperties}).
\par
\begin{figure}[h]
 \centering
  \includegraphics[width=0.50\textwidth]{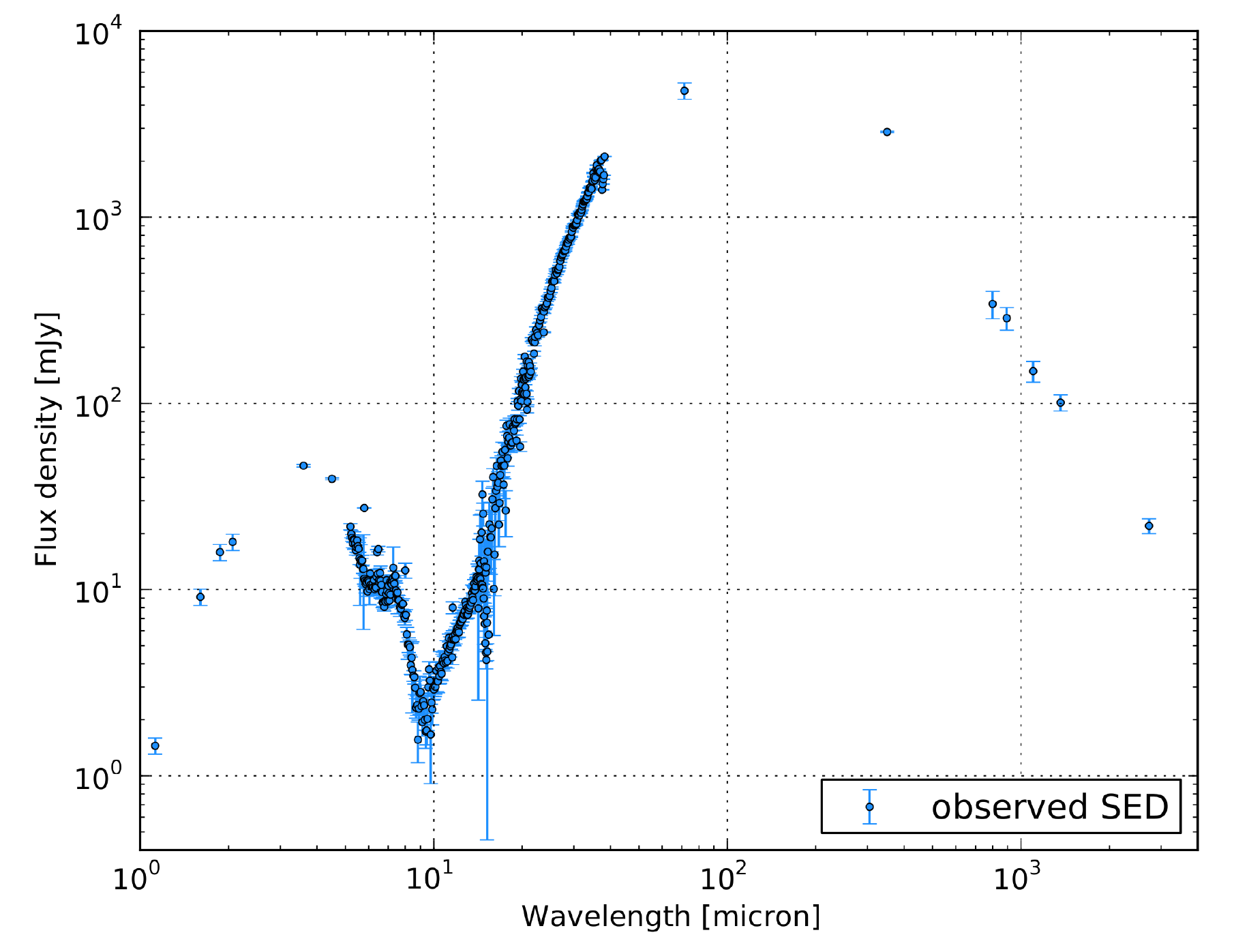}
\caption{Spectral energy distribution of the Butterfly Star.}
\label{Fig:SED}
\end{figure}
\begin{figure}[h]
 \centering
  \includegraphics[width=0.50\textwidth]{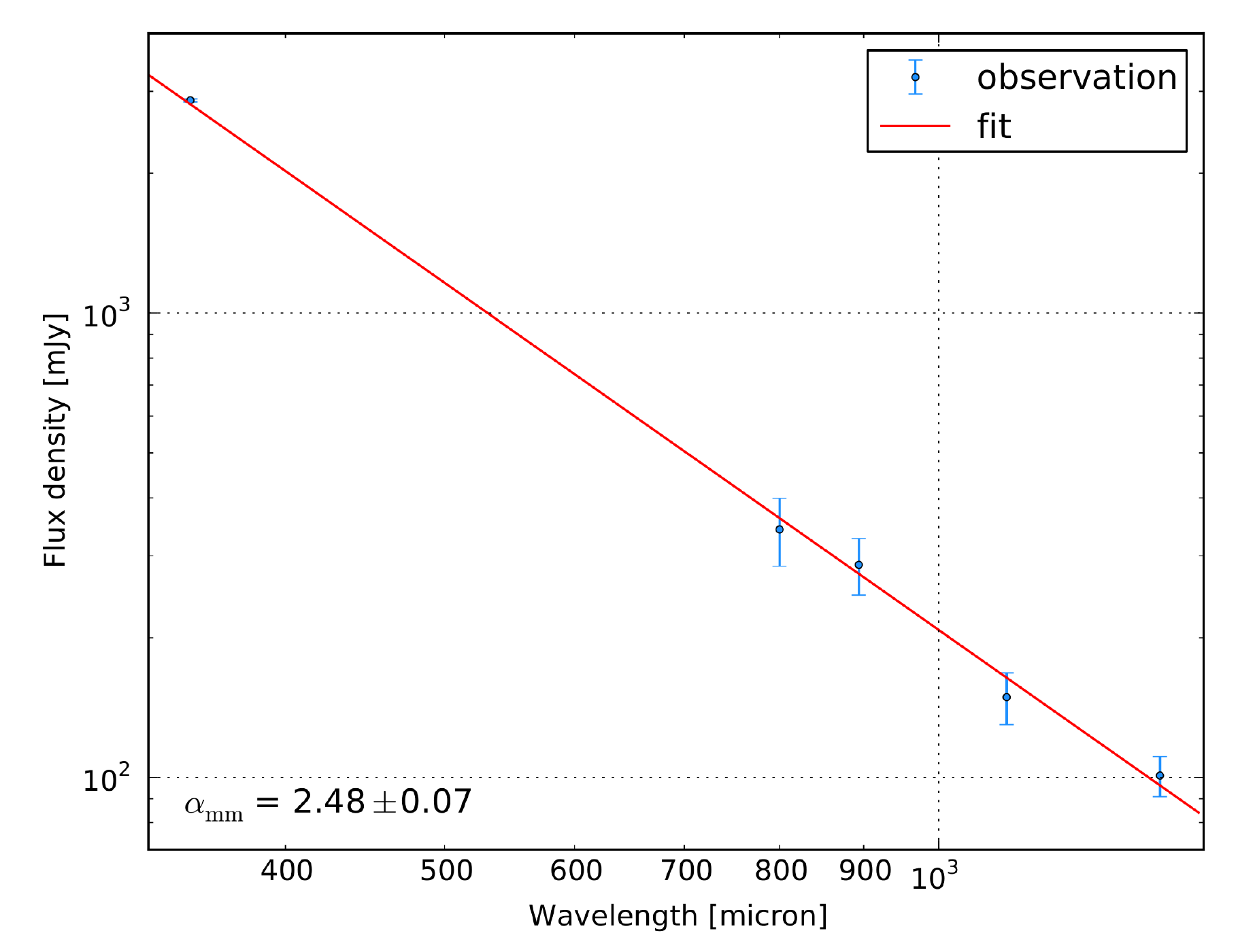}
\caption{Spectral index $\alpha_{\rm mm}$ of the Butterfly Star. For details see Sect.\,\ref{subsection:basis_sed}.}
\label{Fig:spectral_index}
\end{figure}

Figure~\ref{Fig:SED} shows the complete SED of the Butterfly Star consisting of the data shown in Table~\ref{Tab:fluxes}, the SST/IRS spectrum ranging from 5\,\micron\ to 38\,\micron, as well as HST/NICMOS, SST/IRAC, and WISE data points. The SED of our target is typical for a Class I YSO. It peaks close to 100\,\micron\ and shows a 10\,\micron\ absorption feature. This silicate feature and the steep slope of the IRS spectrum beyond 10\,\micron\ require a high inclination angle which is consistent with the spatially resolved NIR images. In Fig.~\ref{Fig:spectral_index} the fit to the (sub)mm slope of the SED of the Butterfly Star can be seen which yields the millimeter spectral index $\alpha_{\rm mm} = -{\rm d}\log(F_{\lambda}) / {\rm d}\log(\lambda) = 2.48\pm0.07$. This value is significantly smaller than that found for small ISM-sized dust grains ($\sim\!3.7$), indicating grain growth in the disk \citep[see, e.g.,][]{Natta07}.
\begin{table}[h]
\caption{Photometric data points for the Butterfly Star used in the modeling}
\label{Tab:fluxes}
\centering
\begin{tabular}{c c c c c}
\hline\hline
$\lambda$ [\micron]	& Flux density [mJy]	& Instrument	& Reference	\\
\hline
22			& $213\pm9$		& WISE		& (1)	\\ 
24			& $241\pm2$		& SST/MIPS	& (2)	\\ 
71			& $4800\pm480$		& SST/MIPS	& (2)	\\ 
350			& $2869\pm21$		& CSO/SHARC-II	& (3)	\\ 
800			& $342\pm57$		& JCMT/UKT14	& (4)	\\ 
894			& $287\pm40$		& SMA		& (5)	\\ 
1100			& $149\pm19$		& JCMT/UKT14	& (4)	\\ 
1364			& $101\pm10$		& PdBI		& (6)	\\ 
2730			& $22\pm2$		& OVRO		& (7)	\\ 
\hline
\end{tabular}
  \tablebib{(1) WISE ALL-Sky Data Release; (2) \citet{Robitaille07}; (3) \citet{Andrews05}; (4) \citet{UKT14_94}; (5) \citet{Wolf08}; (6) This work; (7) \citet{Wolf03}.}
  \tablefoot{WISE: Wide-Field Infrared Survey Explorer; MIPS: Multiband Imaging Photometer for Spitzer; CSO: Caltech Submillimeter Observatory; SHARC-II: Submillimetre High Angular Resolution Camera; JCMT: James Clerk Maxwell Telescope; OVRO: Owens Valley Radio Observatory.\\
  Apertures: $22\,$\micron:\,$16.5\arcsec$; $24\,$\micron:\,$10\arcsec$; $71\,$\micron:\,$20\arcsec$; $350\,$\micron:\,$30\arcsec$; $800\,$\micron:\,$16\arcsec$; $1100\,$\micron:\,$18.7\arcsec$.}
\end{table}

\par

\section{Model description}
\label{section:modeldescription}

  \subsection{The disk}
  \label{subsection:disk}

\begin{figure}[h]
 \centering
  \includegraphics[width=0.45\textwidth]{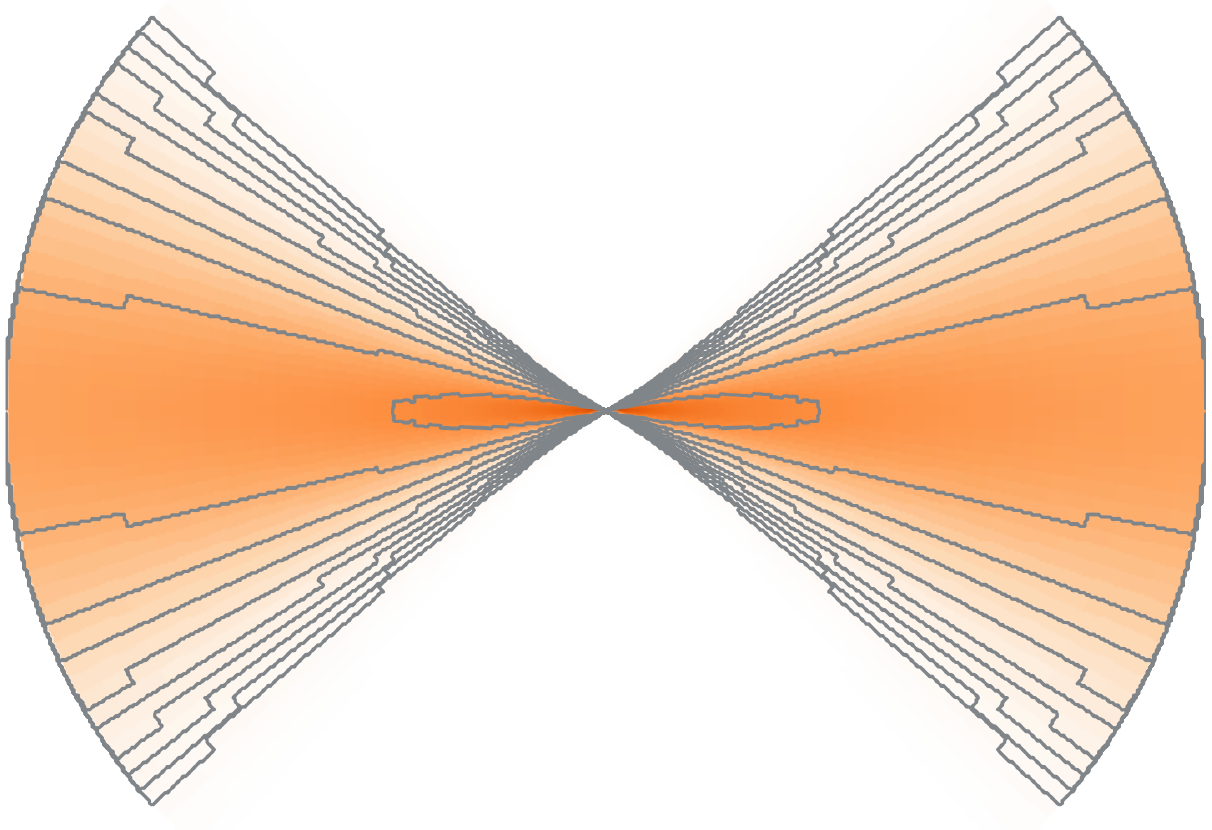}
\caption{Illustration of a flared edge-on oriented disk ($xz$ plane) described by Eq.~\ref{Eq:den_disk}. The contour levels are logarithmic ($\log_{10}$) with outwards decreasing values.}
\label{Fig:den_first}
\end{figure}

Our model for the Butterfly Star consists of a parameterized disk. We describe this disk with a radially and vertically dependent density distribution based on the work of \citet{Shakura73} which can be written as
\begin{gather}
\rho_{\rm disk} = \rho_0 \left(\frac{r_0}{r_{\rm cyl}}\right)^{\alpha} {\rm exp}\left(-\frac{1}{2}\left[\frac{z}{h(r_{\rm cyl})}\right]^2\right).
\label{Eq:den_disk}
\end{gather}
Here, z is the cylindrical coordinate with $z=0$ corresponding to the disk midplane and $r_{\rm cyl}$ is the radial distance from this $z$-axis \citep[see, e.g.,][see Fig.~\ref{Fig:den_first} for illustration]{Wolf03, Stapelfeldt98}. The parameter $\rho_0$ is determined by the total dust mass $m_{\rm dust}$ of the disk. The quantity $r_0$ is a reference radius with $r_0 = 100\,$AU, whereas $h$ is the vertical scale height and a function of $r_{\rm cyl}$
\begin{gather}
h(r_{\rm cyl}) = h_0 \left(\frac{r_{\rm cyl}}{r_0}\right)^{\beta}.
\label{Eq:scale_height}
\end{gather}
The parameters $\alpha$ and $\beta$ describe the radial density profile and the disk flaring, respectively. The disk extends from an inner cylindrical radius $r_{\rm in}$ to an outer one $r_{\rm out}$. Together with the quantity $h_0$, these five parameters are used to adjust the disk structure in order to fit the data. This ansatz has already been successfully applied to other edge-on circumstellar disks, such as HH\,30 \citep{Madlener12}, CB\,26 \citep{Sauter09}, IM\,Lupi \citep{Pinte08}, HV\,Tauri\,C \citep{Stapelfeldt03}, and HK\,Tauri \citep{Stapelfeldt98}. Integrating Eq.~\ref{Eq:den_disk} along the $z$-axis yields the surface density
\begin{gather}
\Sigma\left(r_{\rm cyl}\right) = \Sigma_0\left(\frac{r_{\rm cyl}}{r_0}\right)^{-p} {\hspace{0.4cm}\rm with \hspace{0.4cm}} p = \alpha - \beta.
\label{Eq:surf_den}
\end{gather}
\par

  \subsection{Dust}
  \label{subsection:dust}

In our model we focus on radiative transfer through the dust that dominates the transport of radiation and thermal structure of the disk \citep{ChiangGoldreich97}. The density structure described in Eq.~\ref{Eq:den_disk} is given by the gas in the disk. Because the dust in the disk is coupled to the gas, its density distribution is also described by this equation. The dust grain properties in our model can be divided into three groups: the shape of the dust grains, their chemical composition, and their size distribution.
\par

    \subsubsection{Grain shape}
    \label{subsubsection:grainshape}

We assume the dust grains to be homogeneous, spherical, non-aligned, and non-oriented particles (Mie theory), although they are expected to feature a much more complex and fractal structure. As shown by \citet{Voshchinnikov02}, shape, chemical composition, and size of dust grains cannot be determined separately, but only in combination. Hence, we restrict our model to the simplest but least ambiguous shape possible.
\par

    \subsubsection{Chemical composition}
    \label{subsubsection:grainchemistry}

To model the chemical composition of the dust grains we use the homogeneous mixture of smoothed astronomical silicate and graphite with a mean density of $\rho_{\rm grain} = 2.5\rm\,g\,cm^{-3}$ and optical properties described by \citet{Weingartner01}. Relative abundances of $62.5\%$ astronomical silicate and $37.5\%$ graphite are used \citep{Draine84, Weingartner01}. This grain model has been employed successfully in the modeling of HH\,30 \citep{Madlener12} and CB\,26 \citep{Sauter09}, for example. Given the index of refraction, the optical attributes of a homogeneous sphere can be calculated using Mie theory. We extrapolate the complex refractive indices to a wavelength of $2.7$\,mm to cover all available observational data. This is applicable since both the real and the imaginary part of the refractive index show asymptotic behavior in this wavelength regime. Because graphite is a highly anisotropic material, it is necessary to treat the two possible alignments of the electric field to the crystal axis independently using the so-called $\frac{1}{3} - \frac{2}{3}$ approximation for graphite spheres. That means, if $Q_{\rm ext}$ is the extinction coefficient, then
\begin{gather}
Q_{\rm ext} = \frac{1}{3}Q_{\rm ext}\left(\epsilon_{\parallel}\right) + \frac{2}{3}Q_{\rm ext}\left(\epsilon_{\bot}\right)
\end{gather}
explains the dependence of $Q_{\rm ext}$ on the components of the dielectric tensor ($\epsilon_{\parallel}$ and $\epsilon_{\bot}$) of the electric field parallel and perpendicular to the crystallographic axis. \citet{Draine93} showed that this approximation is sufficiently accurate for extinction curve modeling.
\par

    \subsubsection{Grain sizes}
    \label{subsubsection:grainsizes}

The sizes of the dust grains in our model are distributed according to a power law of the form
\begin{gather}
{\rm d}n\left(a\right) \sim a^{-3.5}\,{\rm d}a {\hspace{0.4cm}\rm with \hspace{0.4cm}} a_{\rm min} < a < a_{\rm max}.
\label{Eq:grain_sizes}
\end{gather}
Here, $a$ is the dust grain radius and $n(a)$ the number of dust grains with a specific radius. For $a_{\rm min} = 5$\,nm and $a_{\rm max} = 250$\,nm this grain-size distribution becomes the commonly-known MRN distribution of the ISM by \citet{Mathis77}. To properly model the different grain sizes in a dust grain mixture, an arbitrary number of separate dust grain sizes within a given interval $[a_{\rm min}$\,:\,$a_{\rm max}]$ has to be considered. \citet{Wolf03b} showed that the observables resulting from radiative transfer (RT) simulations considering each grain species separately are close to those based on weighted mean dust grain parameters of the dust grain mixture. Thus, we use weighted mean values for the optical properties of the dust grain mixture. Moreover, since the separate processes (e.g., grain growth, dust settling, grain-grain interaction, mixing processes) and their mutual influence during the evolution of the circumstellar environment are still rather poorly understood and also to reduce the number of free parameters necessary to describe the dust grain mixture, we assume no spatial dependence of its properties and assume the power law distribution (Eq.~\ref{Eq:grain_sizes}) to be valid throughout the circumstellar disk.
\par

  \subsection{Heating sources}
  \label{subsection:heatingsources}

Stellar heating is the primary heating source for the circumstellar disk. The radiation of the central star heats the dust which then, in turn, re-emits at longer wavelengths. The star, which is represented by a single black body, is characterized in our model by its effective temperature $T_{\star}$ and bolometric luminosity $L_{\star}$. Owing to obscuration by the circumstellar disk, the illuminating source of the system cannot be observed directly, leading to weak constraints on these two parameters from observation.
\par
Viscous heating of the disk is neglected and we ignore accretion and turbulent processes in the disk.
\par

\section{Means of modeling}
\label{section:modeling}

The goal of the modeling process is to find a coherent model for the circumstellar disk of the Butterfly Star explaining all the described observational data. Using continuum RT simulations we compute observables from the model that are then compared to the observed data to find the best-fit model. In this section we describe the means of modeling in detail.
\par

  \subsection{Radiative transfer}
  \label{subsection:radiativetransfer}

For our continuum RT simulations we use the program \texttt{MC3D} \citep{Wolf99, Wolf03MC3D}. It is based on the Monte-Carlo method and solves the continuum radiative transfer problem self-consistently. It makes use of the temperature correction technique described by \citet{Bjorkman01}, the absorption concept introduced by \citet{Lucy99}, and the enforced scattering scheme proposed by \citet{Cashwell59}. The optical properties of the dust grains (scattering, extinction and absorption cross sections, scattering phase function) and their interaction with the radiation field is calculated using Mie theory. Multiple and anisotropic scattering is considered.
\par
In order to derive a spatially resolved dust temperature distribution, the model space has to be subdivided into volume elements inside which a constant temperature is assumed. Both the symmetry of the density distribution and the density gradient distribution have to be taken into account. We use a spherical model space, centered on the illuminating star and an equidistant subdivision of the model in $\theta$-direction, while a logarithmic radial subdivision is applied to resolve the temperature gradient at the very dense inner region of the disk.
\par
The RT is simulated at 107 wavelengths; 100 of them are logarithmically distributed in the wavelength range [0.05\,\micron, 2000\,\micron] and the remaining 7 wavelengths are distributed in the range [894\,\micron, 3000\,\micron]. The quantities we derive with \texttt{MC3D} from the model are
\begin{enumerate}
  \item (Sub)mm maps at 894\,\micron\ and 1.3\,mm;
  \item NIR images at 1.13\,\micron, 1.60\,\micron, 1.87\,\micron, and 2.07\,\micron; and
  \item The SED.
\end{enumerate}
\par

  \subsection{Model fitting properties}
  \label{subsection:modelfittingproperties}

The simulated (sub)mm maps are convolved with the point spread function (PSF) that is described by an elliptical Gaussian function based on the beam major and minor axis as well as the beam position angle from the corresponding observation (see Sect.\,\ref{section:obs_datared}). To account for the rotation of the observed maps, the beam position angle of the PSF is adjusted by the same angle. For the simulated NIR images we use the corresponding PSF obtained with the PSF modeling tool Tiny\,Tim\,v.7.4 \citep{Krist11, Krist04} for the convolution. Additionally, the convolved simulated maps have the same pixel scale as the observed data. This way we ensure that the simulations are comparable to the observations.
\par
There are primary features that we want to reproduce with our model. These also determine the criteria for the best-fit model. For the SED we aim to reproduce the thermal re-emission spectrum over almost three orders of magnitude from the MIR down to the mm wavelength regime. For resolved images the issue is not as simple as for the SED. Our model is rotationally symmetric and thus does not account for any related asymmetry that is seen in the observations. We consider the maps of the disk to be two different groups, the (sub)mm maps and the NIR maps.
\begin{itemize}
  \item[$\bullet$] \textit{Maps in the (sub)mm:} Although the two (sub)mm maps are simply structured they provide two decisive features that constrain our model and that we want to reproduce, the peak flux density and the spatial brightness distribution (mainly the radial brightness distribution with constraints for the vertical extent). The $uv$ data is, in general, used to analyze the (sub)mm data because this avoids the non-linear deconvolution step. Here, as the source is quite strong and barely resolved in one direction, the deconvolution is straightforward and we use image plane comparison.
  \item[$\bullet$] \textit{Maps in the NIR:} The four images in the NIR show more structures and details than the (sub)mm maps. In addition to the disk, which appears as a dark lane, a very complex and highly-structured envelope surrounds the disk. It is not an objective of this study to reproduce this envelope with its wavelength-dependent morphology, so we restrict ourselves to reproducing the width of the dust lane and, consequently, the wavelength dependence of the width of the dust lane. This is the main information provided by the NIR images because it constrains the vertical opacity structure of the disk.
\end{itemize}
\par

  \subsection{Quality of the fit}
  \label{subsection:quality_fit}

For each comparison between model and observation on the above points, we determine a goodness of fit value $\xi_k^2$ that characterizes how well the model fit matches the observational data and so the best-fit model is represented by the smallest goodness of fit value of the investigated parameter space.
\par
\textit{SED:} The goodness of fit of the SED $\xi_{\rm SED}^2$ is given by
\begin{gather}
\xi_{\rm SED}^2 = \frac{1}{N}\sum_{i=1}^N\left(g_i\,\frac{(\mu_i-\omega_i)^2}{\sigma_i^2}\right).
\end{gather}
Here, $N$ is the number of observed wavelengths, $\omega_i$ are the observed flux densities and $\mu_i$ the synthetic flux densities at these wavelengths. The observational uncertainties are considered by $\sigma_i$. As the re-emission part of the SED of the Butterfly Star, which we want to fit, is better sampled in the wavelength range from 12\,\micron\ to 38\,\micron\ thanks to the SST/IRS spectrum than at longer wavelengths up to 2.7\,mm, we include weighting factors $g_i$ in the calculation of the $\xi_{\rm SED}^2$. They are choosen in such a way to ensure that the better and the worse sampled parts of the SED are equally considered in the $\xi_{\rm SED}^2$ calculation, i.e., that the flux measurements ranging from 12\,\micron\ to 38\,\micron\ have in total the same weighting as the data points in the range from 70\,\micron\ to 2730\,\micron\ altogether. A natural weighting fit would rely excessively on a narrow range of wavelengths and could possibly miss gross features which have a much lower S/N. By using an adjusted weighting, we ensure that we do not sacrifice the overall agreement for tiny details in the SED.
\par
\textit{Maps:} For every map the goodness of fit $\xi_{\rm map}^2$ is calculated by
\begin{gather}
\xi_{\rm map}^2 = \frac{\left(\mu-\omega\right)^2}{\sigma^2},
\label{Eq:chi2_map}
\end{gather}
where $\mu$ is the modeled data, $\omega$ the observed data, and $\sigma$ the observational uncertainty. Therefore, for the (sub)mm maps every pixel $j$ whose flux density in the observed data is larger than the background noise in a square box with a side length according to the diameter of the object and centered on the object is taken into account, which lets us rewrite Eq.~\ref{Eq:chi2_map} as
\begin{gather}
\xi_{\rm (sub)mm\,map}^2 = \frac{\frac{1}{N}\sum_{j=1}^N\left(\mu_j-\omega_j\right)^2}{\sigma^2}.
\end{gather}
In this case, $N$ is the number of pixels taken into account in this box and $\sigma$ represents the background noise of the observed map. Instead of a map-based goodness of fit there is also the possibility of choosing an approach where only the profile along the disk axis is taken into account. However, this method would not allow for any direct constraints on the disk scale height. Therefore, we decided to choose the approach described above.
\par
To get a total goodness of fit $\xi_{\rm total}^2$ that characterizes a model fit we combine all $\xi_k^2$
\begin{gather}
\xi_{\rm total}^2 = \frac{1}{N}\sum_{k=1}^N \xi_k^2.
\label{Eq:chi2_total}
\end{gather}
Based on the $\xi_{\rm total}^2$ we get from Eq.~\ref{Eq:chi2_total}, we give our modeling errors, i.e., constraints on the model parameters, as the range where we can alter the parameter values without changing $\xi_{\rm total}^2$ by more than $10\%$. Allowing for a larger variation of $\xi_{\rm total}^2$ than $10\%$ gives generally worse results.
\par

  \subsection{Parameter space study}
  \label{subsection:parameterspacestudy}

The modeling process is divided into two steps. We simultaneously fit the (sub)mm data for the first time and figure out if it is possible to find a model fit that reproduces these data using the described model (Sect.\,\ref{section:modeldescription}) with a single dust grain-size distribution. Based on the results  of this first modeling step (see Sect.\,\ref{subsection:firstmodelingstep_results}) we modify our model setup and also include the SED and the NIR images in the fitting process.
\par
On the basis of the model described in the previous sections, the parameter space we deal with is  defined by several free parameters. These adjustable model parameters are used to reproduce the key features of our observations as described in Sect.\,\ref{subsection:modelfittingproperties}. The volume of the parameter space consists of six and nine free parameters in the first and second modeling step, respectively. For our study, the selection and range of these parameters is based on modeling of other similar objects and previous modeling efforts \citep[see, e.g.,][]{Sauter09, Wolf03}. To constrain the model parameters as strongly as possible and to find the parameter set of the best-fit model we use a grid search based method.
\par

\section{First modeling step}
\label{section:firstmodelingstep}

We first consider a disk model with no spatial variations of the dust grain properties.
\par

  \subsection{Model Parameters}
  \label{subsection:firstmodelingstep_modelparameters}

The six adjustable model parameters we use in the first step are the following:
\begin{itemize}
  \item[$\bullet$] The exponents $\alpha$ and $\beta$ that describe the radial density profile of the disk and the disk flaring, respectively (see Eqs.~\ref{Eq:den_disk} and \ref{Eq:scale_height}). In our modeling, both parameters are treated as independent quantities. We vary $\alpha$ in the range of [1.0, 4.0] and $\beta$ in the range of [1.0, 2.0].
  
  \item[$\bullet$] The scaling factor $h_0$ of the disk's scale height (see Eq.~\ref{Eq:scale_height}). Because the value of the reference radius $r_0$ is $100$\,AU, $h_0$ equals the scale height at a radial distance from the star of $r_{\rm cyl} = 100$\,AU ($h_0 = h({\rm 100\,AU})$). We consider $h_0$ in the range of [1.0, 25.0]\,AU.
  
  \item[$\bullet$] The dust mass $m_{\rm dust}$ of the disk. The dust mass range we probe is [$1.0\times10^{-5}$, $1.0\times10^{-2}$]\,${\rm M_{\odot}}$.

  \item[$\bullet$] The inner disk radius $r_{\rm in}$. Inner holes in protoplanetary disks are a common prediction of theories of disk evolution and are observed in various circumstellar disks \citep[e.g.,][]{Alexander09, Andrews11, Graefe11, Espaillat12}. The inner radius is varied in the range of [0.1, 50.0]\,AU. The lower limit is approximately the dust sublimation radius.

  \item[$\bullet$] The maximum grain radius $a_{\rm max}$. Grain growth is a major issue for protoplanetary disks as it is the first step toward the formation of planets. For this upper radius of the grain-size distribution we consider nine different values: [0.25, 1.0, 10, 30, 50, 75, 100, 300, 1000]\,\micron.
\end{itemize}
For the minimum grain radius $a_{\rm min}$ we use 5\,nm for both modeling steps. The outer radius of the disk $r_{\rm out}$ is fixed at 300\,AU for both steps based on the (sub)mm observations and the modeling results from \citet{Wolf03}. On the basis of the mass of the star of $\sim\!1.7\,{\rm M_{\odot}}$ resulting from observations of the gas  kinematics (Dutrey et al., in prep.) and pre-main-sequence evolutionary tracks \citep[e.g.,][]{Hillenbrand04}, we found that $L_{\star}=5\,{\rm L_{\odot}}$ and $T_{\star}=4500\,$K characterize the central star best. These values are compatible with the radiative transfer simulations and are used in the whole modeling. According to the precisely determined inclination $i$ of the circumstellar disk of the Butterfly Star of $90\degr\pm3\degr$ (Sect.\,\ref{section:introduction}), we fix the inclination at $90\degr$ in the first modeling step. Because of its location in the Taurus-Auriga molecular cloud complex, a distance $d$ of the object of 140\,pc is adopted throughout the modeling process. An illustration of the density structure of a model fit from step one is seen in Fig.~\ref{Fig:den_first}. Table~\ref{Tab:fixed_params} shows an overview of the fixed parameters used in the modeling process.
\par

  \subsection{Results}
  \label{subsection:firstmodelingstep_results}

\hvFloat[floatPos=!htb, capWidth=h, capPos=r, capAngle=90, objectAngle=90, capVPos=c, objectPos=c]
	{figure}
	{\includegraphics[width=1.3\textwidth]{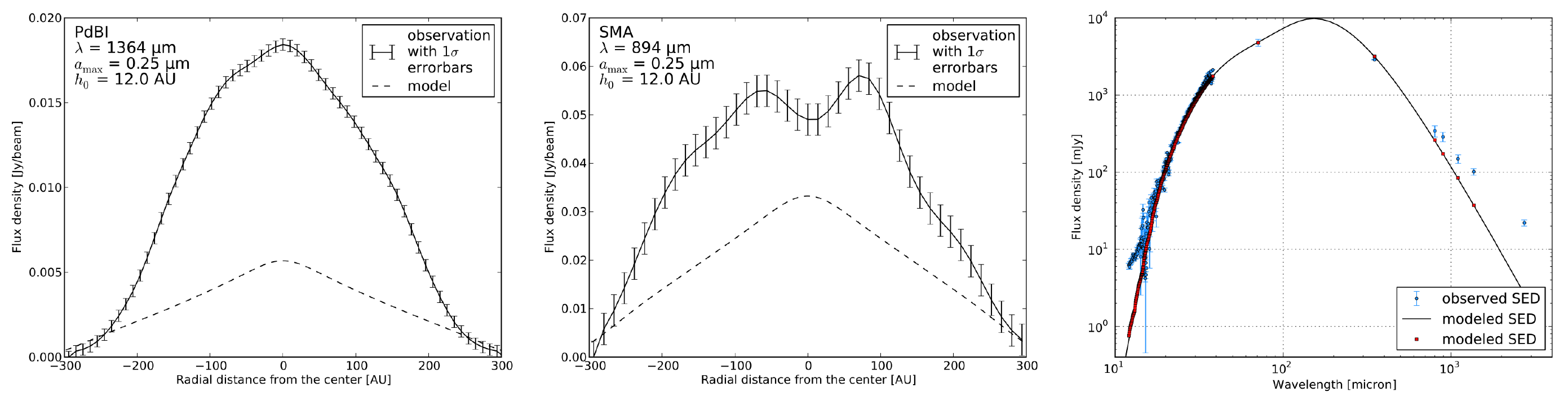}}
	{Radial brightness profiles of the (sub)mm observations and model fits (left, middle) and the SED (right). The plots show the best-fit model of the first step using a dust grain-size distribution with $a_{\rm max} = 0.25$\,\micron. For details see Sect.\,\ref{subsection:firstmodelingstep_results}.}
	{Fig:first_mrn}
\hvFloat[floatPos=!htb, capWidth=h, capPos=r, capAngle=90, objectAngle=90, capVPos=c, objectPos=c]
	{figure}
	{\includegraphics[width=1.3\textwidth]{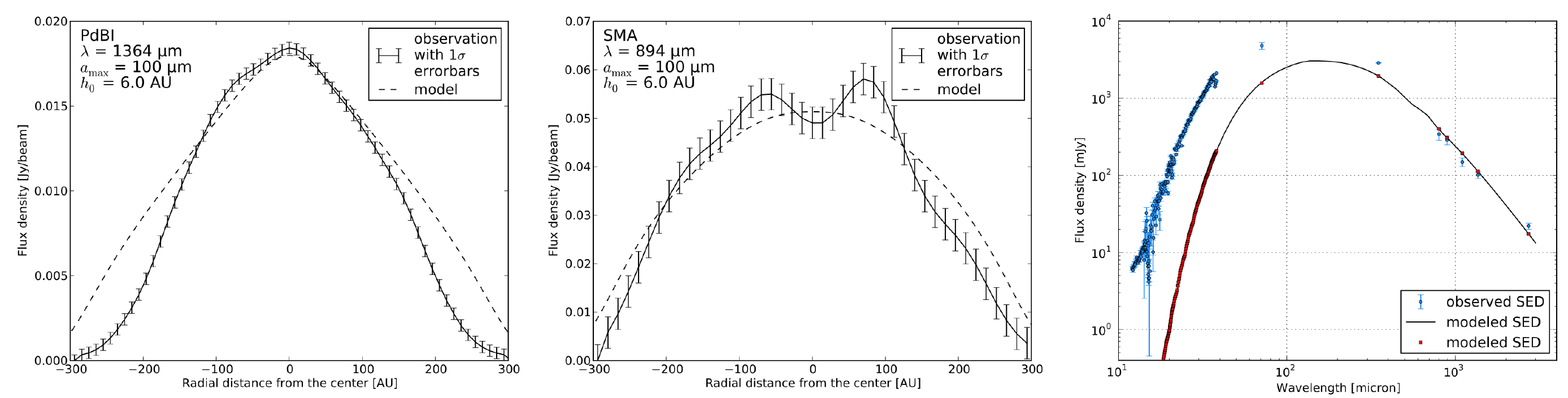}}
	{Radial brightness profiles of the (sub)mm observations and model fits (left, middle) and the SED (right). The plots show the best-fit model of the first step using a dust grain-size distribution with $a_{\rm max} = 100$\,\micron. For details see Sect.\,\ref{subsection:firstmodelingstep_results}.}
	{Fig:first_100micron}
The results presented in this subsection are based on the model and applied assumptions outlined in Sects.\,\ref{section:modeldescription} and \ref{section:modeling}.
\par
Using the described modeling setup of the first step, which is mainly characterized by a single dust grain-size distribution, it is not possible to find a coherent model that properly reproduces the (sub)mm data. Figures~\ref{Fig:first_mrn} and \ref{Fig:first_100micron} show the best-fit results for a grain-size distribution with small ISM-sized dust grains ($a_{\rm max} = 0.25$\,\micron) and larger dust grains ($a_{\rm max} = 100$\,\micron), that yield the motivation for the second modeling step. Dust grains with sizes comparable to those found in the ISM ($a_{\rm max} = 0.25\,$\micron) are neither capable of reproducing the brightness structure of the (sub)mm maps nor do they allow us to fit the shallow mm slope of the SED (Fig.~\ref{Fig:first_mrn}, see also Fig.~\ref{Fig:spectral_index} and Sect.\,\ref{subsection:basis_sed}). The radial brightness distributions of the (sub)mm data in Fig.~\ref{Fig:first_mrn} show an insufficient flux density at almost all regions in the disk. Only in the outermost disk regions the small dust grains yield a flux density that is in agreement with the observation. From the SED in Fig.~\ref{Fig:first_mrn} it can be seen that the MIR and far-infrared (FIR) part is reproduced using ISM-sized dust grains, but that it is not possible to account for the (sub)mm part of the SED. These results strongly suggest the need for larger dust grains in the disk, particularly in the region close to the central star.
\par
We find that the brightness structure of the (sub)mm maps is reproduced best using a dust grain-size distribution with $a_{\rm max} = 100\,$\micron\ (Fig.~\ref{Fig:first_100micron}). Furthermore, the large grains fit the (sub)mm part of the SED better than the small ones while particles significantly smaller or larger than $a_{\rm max} = 100\,$\micron\ are inapplicable. In contrast, it is not possible to account for the IR part of the SED using the large dust grains. Moreover, they yield a flux density that is too high, especially in the outer disk regions as can be seen in the (sub)mm radial brightness profiles (Fig.~\ref{Fig:first_100micron}). This indicates that small dust grains, especially in the outer parts of the disk, are still needed in the model.
\par
Nevertheless, with no combination of the used variable parameters it is possible to account for the proper shape of the radial brightness profiles. We conclude, that to properly fit all the observational data and to find a coherent model that explains all these data, we get from the first modeling step that we need large dust grains with $a_{\rm max} \sim 100$\,\micron\ as well as small ISM-sized dust grains at the same time in the disk model. Additionally, from the disk scale height $h({\rm 100\,AU})$ that amounts to 6\,AU and 12\,AU using the small and the large dust grains, respectively, and from the described brightness profiles (Figs.~\ref{Fig:first_mrn} and \ref{Fig:first_100micron}), we conclude that we need the large grains to be located around the disk midplane with a small vertical extent and concentrated toward the central star.
\par
In addition, no need for a hole in the inner disk is found from the modeling. As a consequence we fixed the inner disk radius in the second modeling step at 0.1\,AU which is approximately the dust sublimation radius.
\par
\begin{table}[!htb]
  \caption{Overview of the fixed parameters.}
  \label{Tab:fixed_params}
  \centering
  \begin{tabular}{l c c c}
    \hline\hline
    Parameter					&	Value	\\
    \hline
    $a_{\rm min}$ [nm]				&	5	\\
    $r_{\rm in}$ [AU]				&	0.1	\\
    $r_{\rm out}$ [AU]				&	300	\\
    $L_{\star}$ [${\rm L_{\odot}}$]		&	5	\\
    $T_{\star}$ [K]				&	4500	\\
    $d$ [pc]					&	140	\\
    \hline
  \end{tabular}
  \tablefoot{The inner disk radius $r_{\rm in}$ is only fixed in the second modeling step.}
\end{table}

\section{Second modeling step}
\label{section:secondmodelingstep}

The results of the first modeling step show that it is not possible to properly fit the (sub)mm data at the same time and that at least two dust grain-size distributions are necessary, with larger dust grains concentrated around the disk midplane.
\par

  \subsection{Model parameters}
  \label{subsection:secondmodelingstep_modelparameters}

\begin{figure}[!htb]
 \centering
  \vspace{0.25cm}
  \includegraphics[width=0.45\textwidth]{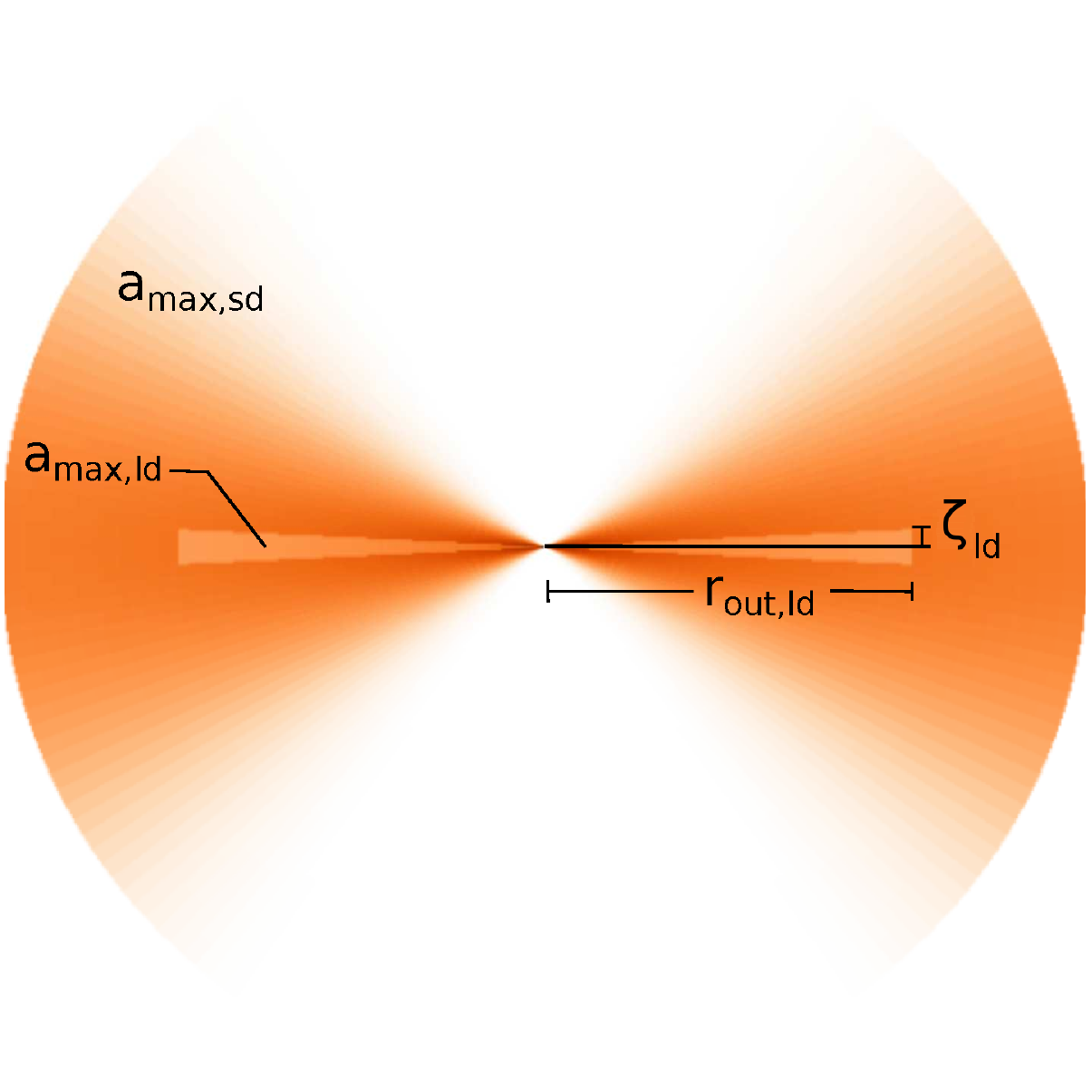}
  \vspace{0.25cm}
\caption{Illustration of a flared edge-on oriented disk ($xz$ plane) using two different grain-size distributions. For details see Sect.\,\ref{subsection:secondmodelingstep_modelparameters}.}
\label{Fig:den_second}
\end{figure}

Owing to the modification of our model setup based on the results of the first modeling step (Sect.\,\ref{subsection:firstmodelingstep_results}), we use a set of nine adjustable parameters in the second step instead of six. In contrast to the first step, we consider two different dust grain-size distributions in the disk model. Their location in the disk can be seen in Fig.~\ref{Fig:den_second}. They differ in the maximum grain size with $a_{\rm max,ld} > a_{\rm max,sd}$ (ld: large dust, sd: small dust). The parameter $a_{\rm max,sd}$ is set to 0.25\,\micron\ which corresponds to the dust grain-size distribution found in the ISM \citep{Mathis77} and that we expect in the outer regions and upper layers of the circumstellar disk of the Butterfly Star \citep{Wolf03}. Because it is not necessary to vary the inner disk radius $r_{\rm in}$ we fix it at 0.1\,AU. The minimum grain radius $a_{\rm min}$ for both grain-size distributions, the outer radius of the disk $r_{\rm out}$, the distance $d$ of the object, $L_{\star}$, and $T_{\star}$ are the same as in the first modeling step (see also Table~\ref{Tab:fixed_params}). The parameters $\alpha$, $\beta$, $h_0$, and $m_{\rm dust}$ are used again as free parameters. In addition, we use the following adjustable parameters in the second step of our modeling:
\begin{itemize}
  \item[$\bullet$] The two quantities $\zeta_{\rm ld}$ and $r_{\rm out,ld}$ that describe the vertical and radial extent of the region where the grain-size distribution with the larger dust grains is located (see Fig.~\ref{Fig:den_second}). The parameter $\zeta_{\rm ld}$ is radially dependent in the same way as in Eq.~\ref{Eq:scale_height}. We emphasize that $\zeta_{\rm ld}$ is not a scale height in the sense of $h_0$. In our model the circumstellar disk is described by one density distribution (see Eq.~\ref{Eq:den_disk}). Therefore, the vertical extent described by $\zeta_{\rm ld}$ has a sharp upper limit. The range we consider for $\zeta_{\rm ld}$ (at 100\,AU) is [1.0, 25.0]\,AU and for $r_{\rm out,ld}$ it is [50, 300]\,AU.

  \item[$\bullet$] A scaling factor $sm$ that adjusts the distribution of the dust mass of the disk within the two different dust regions. This parameter is affected by the other free disk parameters, i.e., if the total number of dust grains is changed in either of the two different dust regions by any of the adjustable parameters, the ratio of the dust mass of these two regions is modified although this scaling factor is not varied. 

  \item[$\bullet$] The inclination $i$ of the circumstellar disk. Despite the small uncertainty of the inclination measurement, here we allow the inclination to vary in the range of $90\degr\pm3\degr$.

  \item[$\bullet$] The maximum grain radius of the dust grain-size distribution containing the larger dust grains $a_{\rm max,ld}$. We consider three different values: [50, 100, 150]\,\micron.
\end{itemize}
Using this model setup in the second modeling step we are accounting for grain evolution and its dependence on the radial and vertical position in the circumstellar disk.
\par

  \subsection{Results}
  \label{subsection:secondmodelingstep_results}

\begin{table*}[!htb]
  \caption{Overview of the parameter ranges, best-fit values, and constraints on the model parameters of the second modeling step.}
  \label{Tab:overview}
  \centering
  \begin{tabular}{l c c c c c c}
    \hline\hline
    Parameter				& Minimum value	& Maximum value	& \bf{Best-fit value}	& Lower limit	& Upper Limit	& $\delta$ \\
    \hline
    $\alpha$				& 1.0		& 4.0		& \bf{1.2}		& 1.2		& 1.6		& 0.2	   \\
    $\beta$				& 1.0		& 2.0		& \bf{1.14}		& 1.12		& 1.16		& 0.01	   \\
    $h_0$ [AU]				& 1.0		& 25.0		& \bf{10.0}		& 9.0		& 11.0		& 1.0	   \\
    $\zeta_{\rm ld}$ [AU]			& 1.0		& 25.0		& \bf{6.0}		& 4.0		& 7.0 		& 1.0	   \\
    $r_{\rm out,ld}$ [AU]		& 50		& 300		& \bf{175}		& 170		& 210		& 5.0	   \\
    $sm$				& 0.001		& 0.9		& \bf{0.083}		& 0.068\tablefootmark{a}	& 0.093\tablefootmark{a}	& 0.001\tablefootmark{a}   \\
    $m_{\rm dust}$ [${\rm M_{\odot}}$]	& 0.00001	& 0.01		& \bf{0.0009}		& 0.0008	& 0.001		& 0.0001   \\
    $a_{\rm max,ld}$ [\micron]		& 50		& 150		& \bf{100}		& $>50$		& $<150$	& 50	   \\
    $i$ [$\degr$]				& 87		& 93		& \bf{90}		& 90		& 90		& 1	   \\
    \hline
  \end{tabular}
  \tablefoot{
    In the last column the step width $\delta$ between the given lower and upper limits of each parameter is given.\\
    \tablefoottext{a}{The constraints on the scaling factor $sm$ refer to the best-fit model. As this parameter is affected by the other parameters, no constraints for the whole investigated parameter space are given (see also Sect.\,\ref{subsection:secondmodelingstep_modelparameters}).}
  }
\end{table*}
The results presented in this subsection are based on the model and applied assumptions outlined in Sects.\,\ref{section:modeldescription} and \ref{section:modeling}.
\par
\begin{figure*}[!htb]
 \centering
  \includegraphics[height=0.45\textheight]{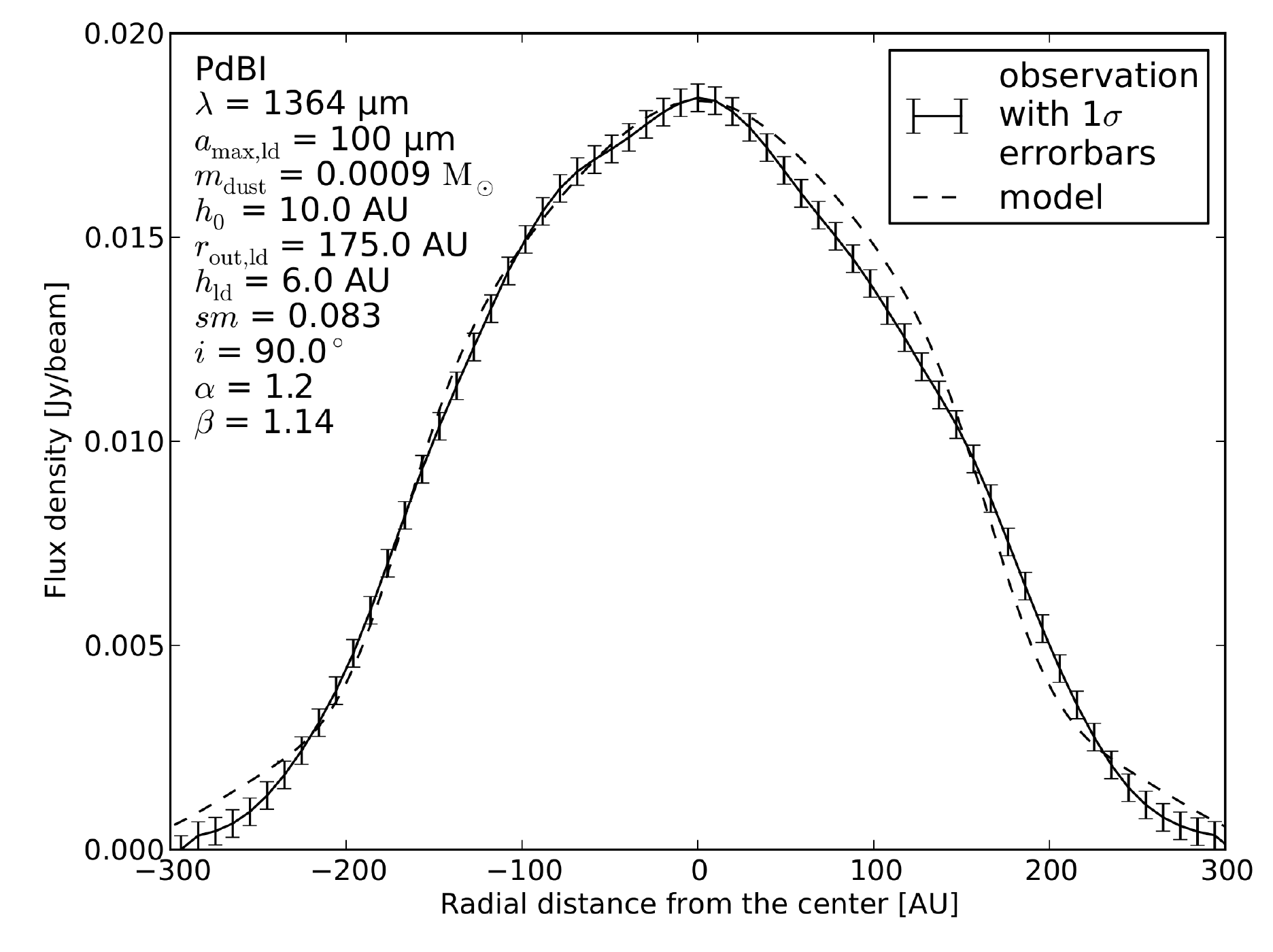}
\caption{Radial brightness profile at 1.3\,mm based on the observation and our best-fit model. For details see Sect.\,\ref{subsection:secondmodelingstep_results}.}
\label{Fig:second_result_pdb}
\end{figure*}
\begin{figure*}[!htb]
 \centering
  \includegraphics[height=0.45\textheight]{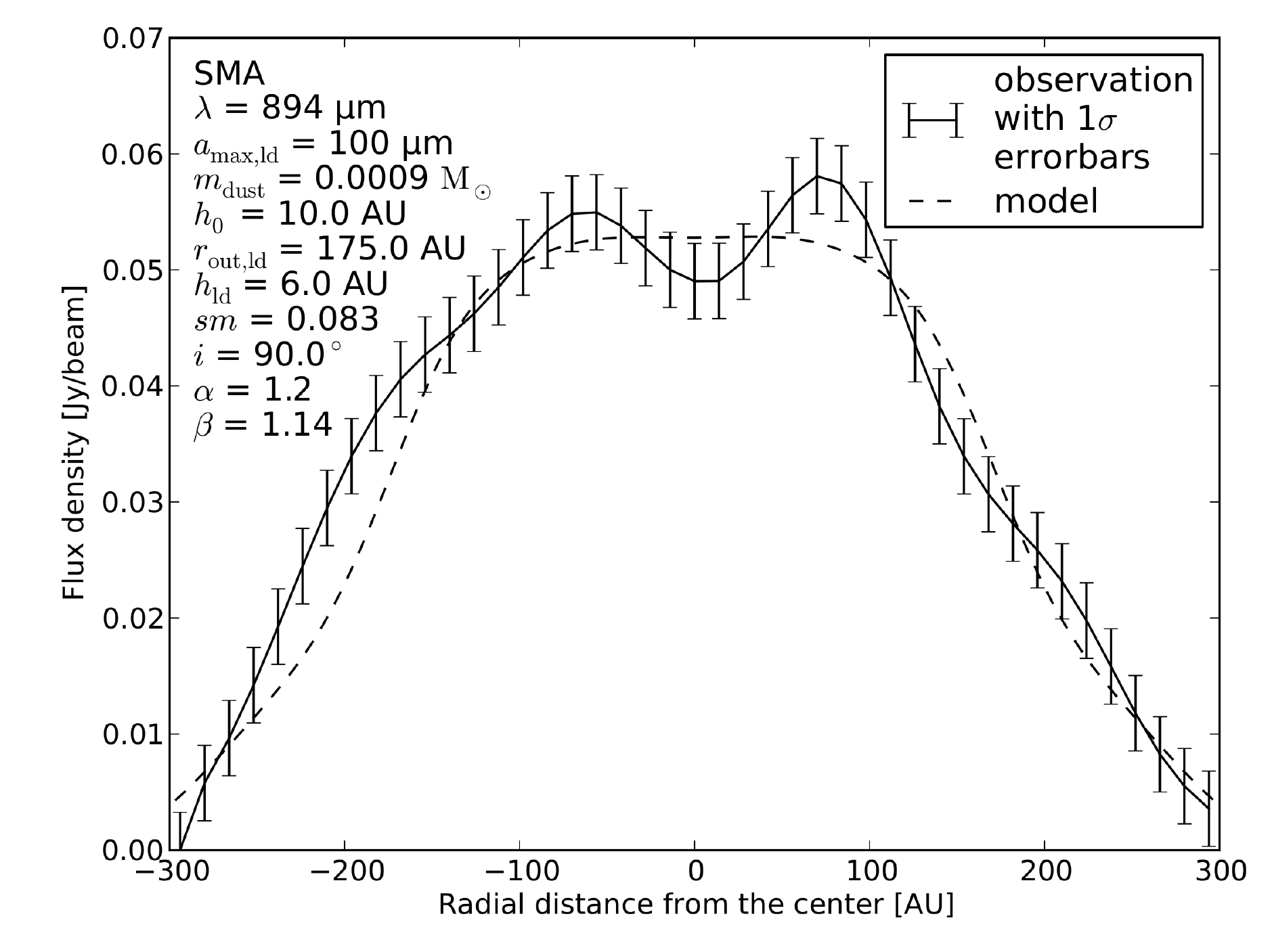}
\caption{Radial brightness profile at 894\,\micron\ based on the observation and our best-fit model. For details see Sect.\,\ref{subsection:secondmodelingstep_results}.}
\label{Fig:second_result_sma}
\end{figure*}
\begin{figure*}[!htb]
 \centering
  \includegraphics[height=0.45\textheight]{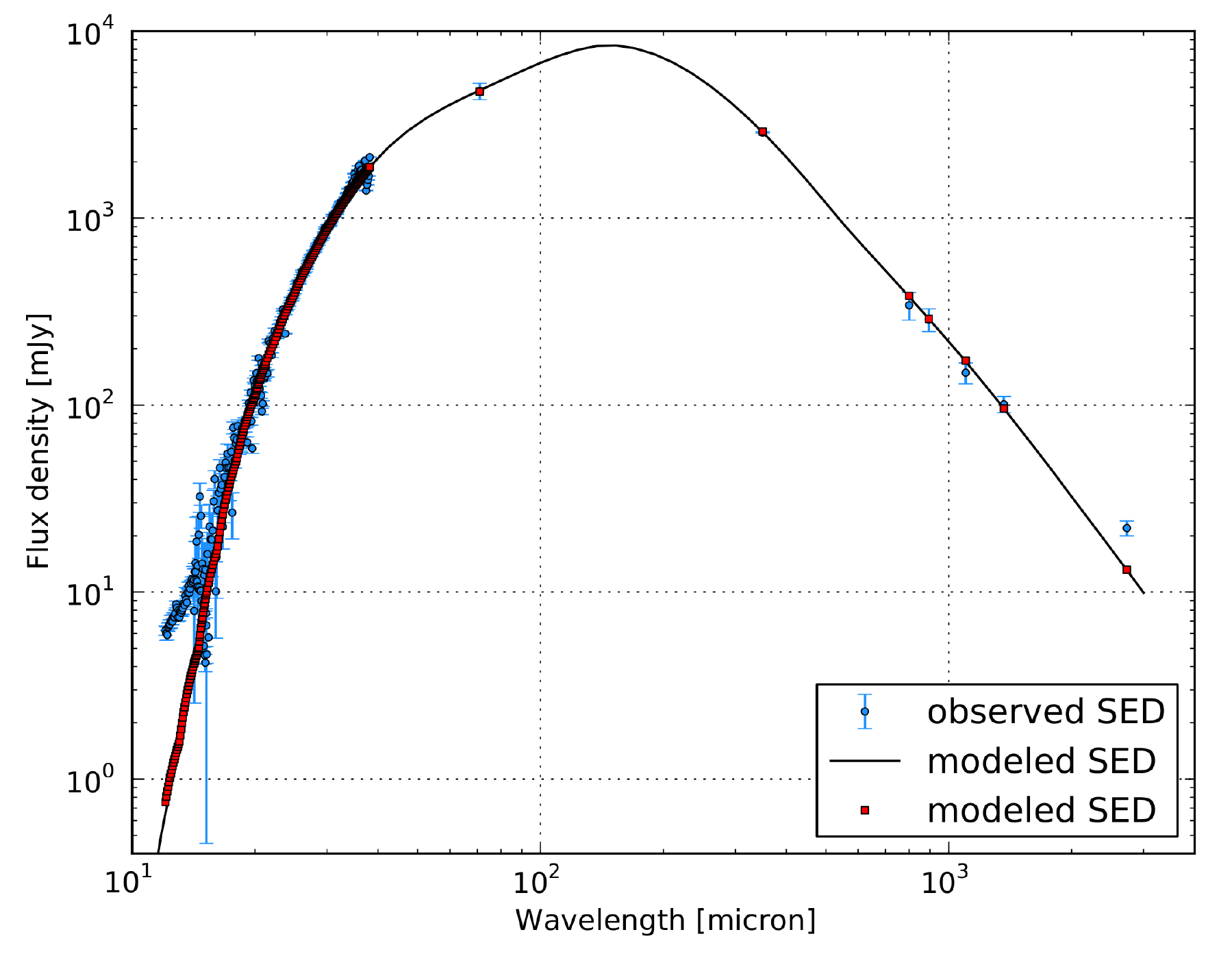}
\caption{SED based on the observations and our best-fit model. For details see Sect.\,\ref{subsection:secondmodelingstep_results}.}
\label{Fig:second_result_sed}
\end{figure*}

With the model setup of the second modeling step, which is mainly characterized by two different dust grain-size distributions, it is possible to find a model fit that explains all the observational data taken into account and therefore a coherent model for the circumstellar disk of the Butterfly Star. The values of the parameters of our best-fit model and the constraints on the model parameters resulting from our multi-wavelength modeling can be found in Table~\ref{Tab:overview}. Only in the small parameter space described by the lower and upper limits of the listed parameters is it possible to find model fits that satisfactorily reproduce all observational data, i.e., $\xi^2_{\rm total}$ does not change by more than 10\,\% compared to the best-fit value. Figures~\ref{Fig:second_result_pdb} and \ref{Fig:second_result_sma} show the radial brightness profiles of the (sub)mm observations overlayed with the corresponding counterparts based on the best-fit model. The optical depth $\tau$ at $894\,$\micron\ amounts to $1.75$ and at $1.3\,$mm to $0.64$. This means that the disk is optically thick at submm wavelength and optically thin at mm wavelength. The effective specific dust opacity is determined to be $2.58\,{\rm cm^2\,g^{-1}}$ and $0.96\,{\rm cm^2\,g^{-1}}$ at $894\,$\micron\ and $1.3\,$mm, respectively (based on the dust model outlined in Sect.\,\ref{subsection:dust}). The observed and modeled re-emission SED is represented in Fig.~\ref{Fig:second_result_sed}. In Table~\ref{Tab:width_dustlane} the width of the dust lane at the point of minimum separation between the two lobes in both the observed and modeled scattered light maps is listed (see also Fig.~\ref{Fig:BS_RGB}). The width of the dust lane stays almost constant across the whole radius of the disk such that small deviations are covered by the listed uncertainties. Overall, our best-fit model is in very good agreement with all considered observations, especially taking into account the range of wavelengths and the variety of observations analyzed in the fitting procedure.
\par
In Fig.~\ref{Fig:second_result_pdb} it can be seen that our model almost perfectly fits the 1.3\,mm observation. The overall accuracy, i.e., how good the modeled profile matches the observed one in terms of the observed background noise, is $1.39\,\sigma_{\rm mm} \pm 0.98\,\sigma_{\rm mm}$. Although the central brightness minimum in the submm observation (Fig.~\ref{Fig:second_result_sma}) is not reproduced by our model we do see a clear flattening of the profile which is definitely caused by an optical depth effect (see also Sect.\,\ref{subsection:minimum_submm}). With an accuracy of $0.96\,\sigma_{\rm submm} \pm 0.77\,\sigma_{\rm submm}$, our model also provides a satisfying fit to the submm observation. Furthermore, the modeled SED is in very good agreement with observations from the MIR over the FIR to the mm regime, meaning that almost all modeled flux densities are within the uncertainties of the observed flux densities (Fig.~\ref{Fig:second_result_sed}). In particular, it reproduces the (sub)mm flux densities and hence the mm spectral index. The width of the dust lane in the modeled NIR images is in agreement with the observed width taking into account the measurement uncertainties (Table~\ref{Tab:width_dustlane}). That means that the four modeled NIR images nicely reproduce the observed width of the dust lane and therefore also its wavelength dependence which was the goal for this wavelength range.
\par
\begin{table}[!htb]
  \caption{Width of the dust lane in the observed NIR maps and those of our best-fit model.}
  \label{Tab:width_dustlane}
  \centering
  \begin{tabular}{c c c c c c}
    \hline\hline
		&	1.12\,\micron	& 1.60\,\micron	& 1.87\,\micron	& 2.07\,\micron	& $\Delta$	\\
    \hline
    Model	&	$1.125\arcsec$	& $0.750\arcsec$	& $0.600\arcsec$	& $0.525\arcsec$	& $\pm0.075\arcsec$	\\
    Observation	&	$1.200\arcsec$	& $0.825\arcsec$	& $0.600\arcsec$	& $0.525\arcsec$	& $\pm0.075\arcsec$	\\
    \hline
  \end{tabular}
  \tablefoot{In the last column the uncertainty $\Delta$ of the measurements is given.}
\end{table}

\section{Discussion}
\label{section:discussion}

  \subsection{Brightness minimum in the SMA observation}
  \label{subsection:minimum_submm}

The central brightness minimum seen in the observed $894$\,\micron\ map was first described in \citet{Wolf08} where two explanations for this minimum are discussed, an inner disk hole and an optical depth effect. When comparing the beam size of the two (sub)mm maps (see Sect.\,\ref{section:obs_datared}) it can be seen that the beam size of the $1.3\,$mm observation is significantly smaller than that of the $894$\,\micron\ map. This means that any disk hole that would be capable of producing such a local brightness minimum in the submm observation would also show up in the 1.3\,mm observation. However, at $1.3\,$mm no sign of a central brightness minimum or even a flattening of the profile was found. Therefore, this rules out the possibility that the local minimum seen in the $894$\,\micron\ map is caused by the lack of emitting dust, i.e., an inner disk hole. Moreover, when comparing both maps and taking the beam sizes into account, we can conclude that the disk of the Butterfly Star is optically thick even at submm wavelength. Thus, the central brightness minimum in the submm observation is produced by an optical depth effect \citep[for details see][]{Wolf08}. This is also supported by the optical depth at both wavelengths (see Sect.\,\ref{subsection:secondmodelingstep_results}) resulting from our best-fit model. We emphasize that the observed brightness minimum has a low significance of $\sim\!1.4\,\sigma$ and $\sim\!1.8\,\sigma$ with respect to the maxima to the left and right of the center, respectively. As our best-fit model shows a clearly flattened profile in the submm that accounts for the optical depth effect and is in very good agreement with the observation (Sect.\,\ref{subsection:secondmodelingstep_results}), it is a satisfying result.
\par

  \subsection{Dust evolution}
  \label{subsection:dustevolution}

We found a coherent model capable of explaining all considered observations. While it was not possible to achieve this goal with a model using a single dust grain-size distribution we succeeded by using two different distributions. Both distributions differ in the size of the maximum particle radius $a_{\rm max}$ and in the spatial distribution in the disk. We found that for $a_{\rm max}$ values of 0.25\,\micron\ and 100\,\micron\ for the smaller and larger dust grain-size distribution, respectively, fit the data best. The grain size we determine for the larger dust in the disk is almost three orders of magnitude larger than upper grain sizes given for the ISM in the literature. This strongly indicates that grain growth is taking place in the circumstellar disk of the Butterfly Star. However, this dust grain size is still one order of magnitude smaller than the maximum grain size derived from modeling other young circumstellar disks, such as for IM\,Lupi \citep{Pinte08}. There, a maximum size of a few mm has been found. In the framework of our model we exclude the strong influence of such large grains on the (sub)mm observations of the disk of the Butterfly Star. Our model clearly shows the need for small ISM-sized as well as larger ($a_{\rm max} \sim 100\,$\micron) dust grains in this disk. In particular, the smaller grains are located in the upper layers and outer region of the disk and the larger ones in the inner disk region. The findings on the grain sizes are only valid in the context of the assumed dust properties (see Sect.\,\ref{subsection:dust}). Assuming a different distribution of the sizes of the dust grains may lead to different results. But what in general can be concluded is that the larger dust grains have a different spatial distribution than the smaller ones.
\par
Comparing the disk scale height of our model and the corresponding vertical extent of the region where the large dust grains are located at 100\,AU yields that the large dust grains are settled toward the disk midplane. Furthermore, comparing the radial extent of the large dust region with the outer disk radius it can be seen that the large dust grains are not distributed over the whole radial extend of the circumstellar disk, but are concentrated toward the star.
\par
We conclude that besides grain growth in the disk of the Butterfly Star, we find evidence for vertical settling and radial segregation of the dust in this disk. These three mechanisms are the key issues in the process of planetary core formation, which start the formation process of planetesimals. Our coherent model not only gives strong constraints on the dust properties in this disk, but we can also provide quantitative constraints on the vertical and radial distribution of the phase with large dust grains for the first time.
\par

  \subsection{Disk mass}
  \label{subsection:diskmass}

Our best-fit model shows that we need a dust mass of $M_{\rm dust}=9\times 10^{-4}\,{\rm M_{\odot}}$ to account for the observed flux densities. This value is well constrained (see Table~\ref{Tab:overview}) and stronger variations of the dust mass would have a significant influence on the re-emission SED, for example. The dust mass of the region where the large grains are distributed is $2.2\times 10^{-4}\,{\rm M_{\odot}}$ and that of the small grains is $6.8\times 10^{-4}\,{\rm M_{\odot}}$. Assuming a typical dust-to-gas ratio of 
\begin{gather}
\frac{M_{\rm dust}}{M_{\rm gas}} \sim \frac{1}{100},
\end{gather}
we get a total disk mass of $\sim\!9\times 10^{-2}\,{\rm M_{\odot}}$ and determine a maximum absorption opacity of $1.97\,{\rm cm^2\,g^{-1}}$ at $1.3\,$mm. This is in good agreement with the results of \citet{Wolf03} and very similar to other YSOs such as IM\,Lupi \citep{Pinte08}, DL\,Tau, GO\,Tau, and HL\,Tau \citep{Andrews07}.
\par
The derived disk mass depends on the adopted assumptions about the chemistry and shape of the dust grains (see Sect.\,\ref{subsection:dust}). Fractal grain shapes and the spatial orientation of every dust grain are not taken into account as this would only introduce further free parameters without providing a qualitative improvement of our understanding of the disk properties and distribution of large grains. The complex shapes of the dust grains would have implications on the light scattering and absorption behavior of the grains \citep[e.g.,][]{Wright87, Lumme94}. Besides that, it also can be thought of dust grains with almost the same absorption cross section as spherical dust grains but with much less mass. Such fluffy particles with a porosity up to 90\% are discussed by \citet{fluffyparticles07}.
\par
Furthermore, we point out that the disk mass is proportional to the grain density $\rho_{\rm grain}$ and the number of dust grains. Within our study we can constrain the disk structure, the grain size, and their number, but not the density of one grain. For our investigation of the circumstellar disk, we used $\rho_{\rm grain} = 2.5\rm\,g\,cm^{-3}$, but a more porous, fractal structure of the dust grains may result in a smaller value. In turn, this will alter our estimate for the disk mass by the same factor. However, our general results on the different spatial distributions of the smaller and larger dust grains are not affected by these assumptions.
\par

  \subsection{Disk structure}
  \label{subsection:diskstructure}

\begin{figure}[h]
 \centering
  \includegraphics[width=0.49\textwidth]{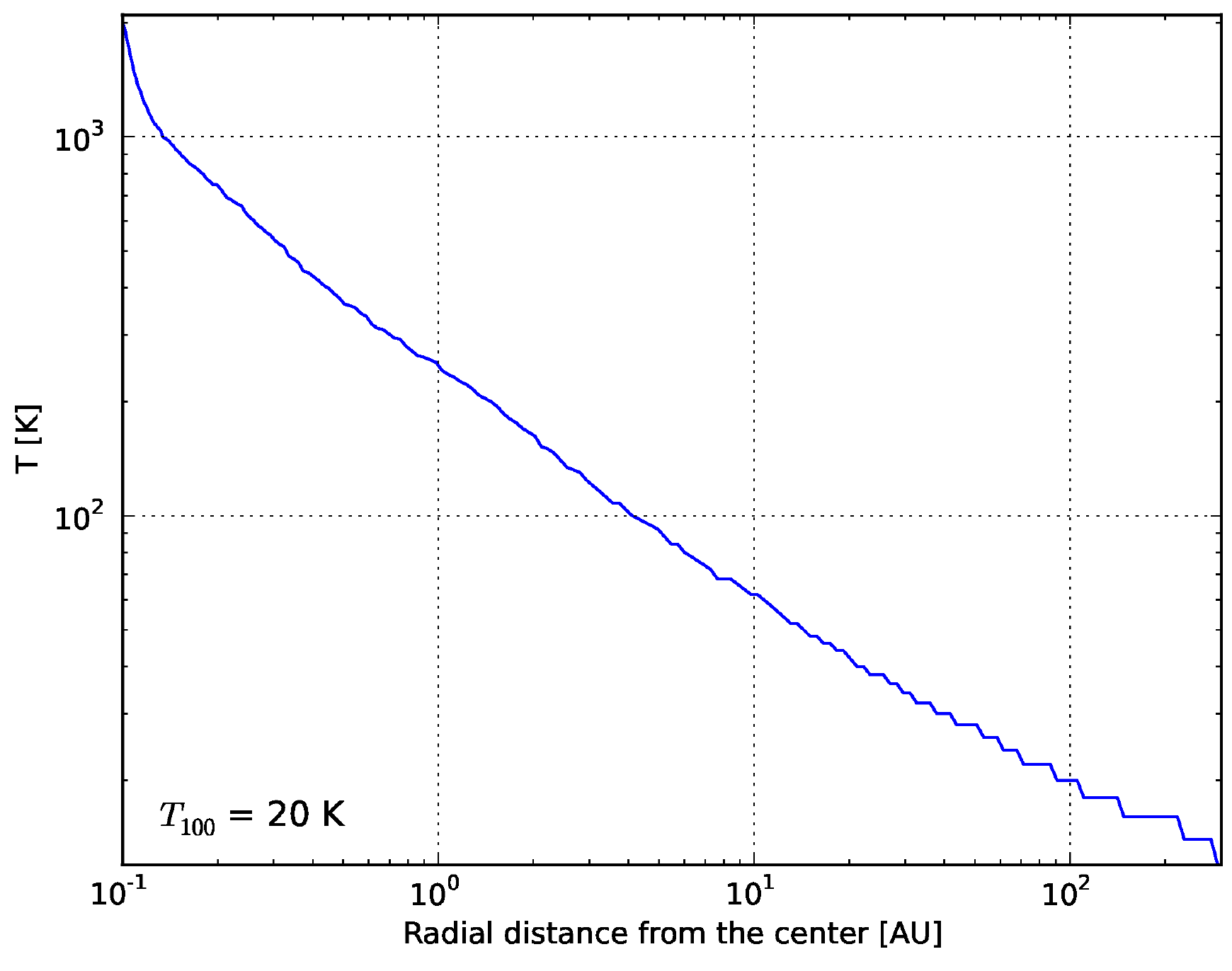}
\caption{Midplane temperature of our best-fit model. $T_{\rm 100}$ is the midplane temperature at 100\,AU.}
\label{Fig:tem_profile}
\end{figure}

Our multi-wavelength modeling allows us to quantitatively constrain most of the geometrical parameters of the circumstellar disk of the Butterfly Star. A flared geometry with a scale height of $\sim\!10\,$AU at a reference radius of 100\,AU is required. The determined flaring index $\beta$ of $\sim\!1.14$ and scale height are comparable to the values for the YSO IM\,Lupi by \citet{Pinte08}. \citet{Sauter09} also found for the circumstellar disk in the Bok globule CB\,26 a scale height of $\sim\!10\,$AU and with $\beta \sim\!1.4$ a slightly larger flaring index. For the circumstellar disk of HH\,30, \citet{Madlener12} found a flaring index comparable to what we find for the disk of the Butterfly Star, but a larger scale height ($h\sim\!15\,$AU). Our best-fit model has a midplane temperature of 20\,K at 100\,AU. Figure~\ref{Fig:tem_profile} shows the calculated midplane temperature as a function of the radial distance from the center. The determined value for $T_{\rm 100}$ is very similar to other YSOs such as IM\,Lupi \citep{Pinte08}, CB\,26 \citep{Sauter09}, DG\,Tau, DG\,Tau-b, HL\,Tau \citep{Guilloteau11}, and HH\,30 \citep{Madlener12}.
\par
The hydrostatic scale height $H$ is derived by
\begin{gather}
H = \sqrt{\frac{k_{\rm B} T(r) r^3}{G M_{\star} \mu m_{\rm p}}},
\end{gather}
where $k_{\rm B}$ is Boltzman's constant, $G$ is the gravitational constant, $\mu = 2.33\,{\rm g\,mol^{-1}}$ \citep{Ruden91} is the mean molecular weight, and $m_{\rm p}$ is the proton mass. This yields a value of $H = 6.9\,$AU at a radius of 100\,AU. To compare this to the vertical extent of the region where the large dust grains are located ($\zeta_{\rm ld}$) an equivalent scale height at a radius of 100\,AU $h_{\rm ld}$ can be determined by
\begin{gather}
h_{\rm ld} = \left(\sqrt{\frac{2}{\pi}}\cdot \int_0^{\zeta_{\rm ld}} z^2\cdot {\rm exp}\left(-\frac{1}{2}\left[\frac{z}{h_0}\right]^2\right) {\rm d}z\right)^\frac{1}{3}.
\end{gather}
For our best-fit model this results in a scale height for the large dust of $h_{\rm ld} = 3.7\,$AU and for the constraints on $h_0$ and $\zeta_{\rm ld}$ in a range of $2.5\,$AU$\le h_{\rm ld}\le 4.3\,$AU. This is substantially smaller than $\zeta_{\rm ld}$ and substantially smaller than the hydrostatic scale height $H$ which strongly indicates that the large dust grains in the circumstellar disk of the Butterfly Star are vertically settled.
\par

    \subsubsection{Surface density}
    \label{subsubsection:surfacedensity}

\begin{figure}[h]
 \centering
  \includegraphics[width=0.49\textwidth]{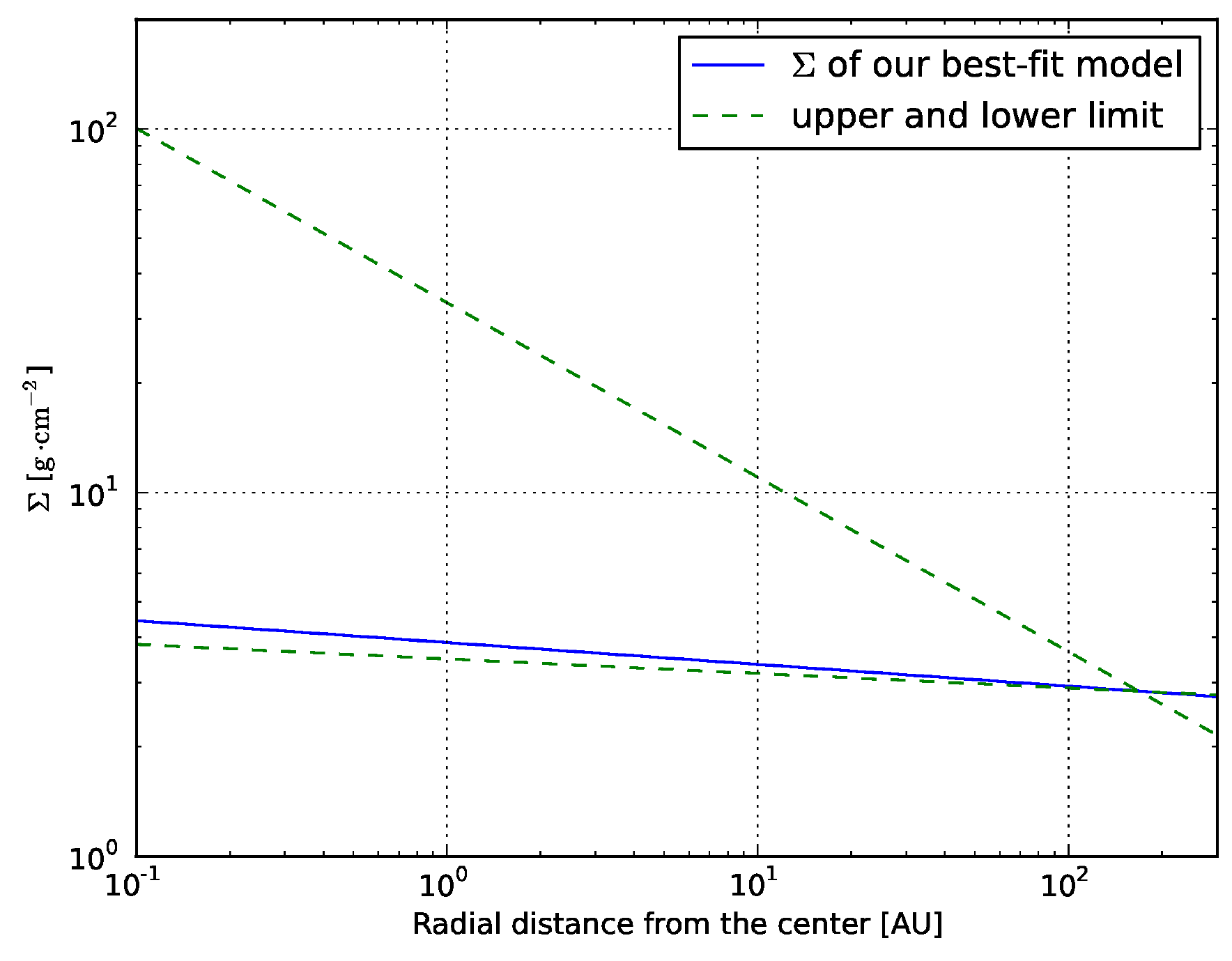}
\caption{Surface density. The upper and lower limit results from the constraints on the model parameters (see Table~\ref{Tab:overview}).}
\label{Fig:surf_den}
\end{figure}

In Fig.~\ref{Fig:surf_den} the derived surface density of our best-fit model for the circumstellar disk of the Butterfly Star as well as the upper and lower limit can be seen. The surface density exponent (see Eq.~\ref{Eq:surf_den}) is found to be in the range of [0.04, 0.48]. These values are lower than those predicted by most theoretical models of disks \citep[$p\sim\!1$;][]{Bell97}. Furthermore, studies of large samples of circumstellar disks in the Taurus-Auriga and Ophiuchus-Scorpius star formation regions by, e.g., \citet{Kitamura02}, \citet{Andrews07}, and \citet{Guilloteau11} show that values of $p$ less than $\sim\!1$ are common, but they are in general larger than $0.5$. Therefore, our result for $p$ is slightly smaller than that found for most other objects. We determine the surface density to be in the range of [3.3, 15.2]\,$\rm\,g\,cm^{-2}$ and [2.9, 3.6]\,$\rm\,g\,cm^{-2}$ at 5\,AU and 100\,AU, respectively, which is consistent with the previously mentioned surveys.
\par

    \subsubsection{Disk stability}
    \label{subsubsection:diskstability}

\begin{figure}[h]
 \centering
  \includegraphics[width=0.49\textwidth]{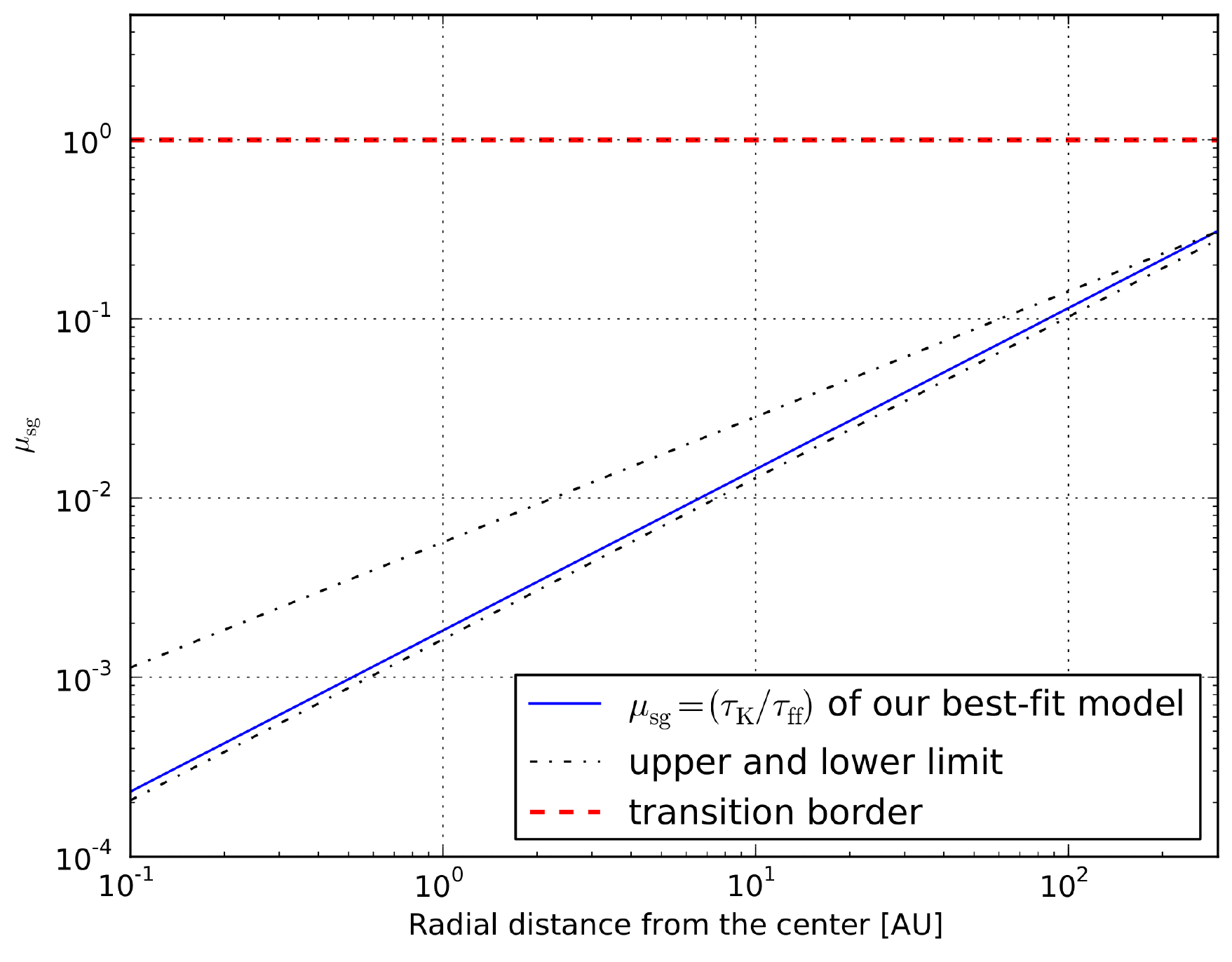}
\caption{Ratio $\mu_{\rm sg}$. The upper and lower limit results from the constraints on the model parameters (see Table~\ref{Tab:overview}). The transition border marks the point where the circumstellar disk becomes self-gravitating.}
\label{Fig:self_grav}
\end{figure}

To estimate if the circumstellar disk of the Butterfly Star is self-gravitating, we determined the ratio $\mu_{\rm sg}$ of the Kepler time $\tau_{\rm K}$ and the local free-fall time $\tau_{\rm ff}$
\begin{gather}
\tau_{\rm K} = \Omega_{\rm K}^{-1} = \sqrt{\frac{r_{\rm cyl}^3}{GM_\star}},
\end{gather}
\begin{gather}
\tau_{\rm ff} \approx \sqrt{\frac{1}{G\rho_{\rm c}}} {\hspace{0.4cm}\rm with \hspace{0.4cm}} \rho_{\rm c} = \rho_0 \left(\frac{r_0}{r_{\rm cyl}}\right)^\alpha,
\end{gather}
\begin{gather}
\mu_{\rm sg} = \frac{\tau_{\rm K}}{\tau_{\rm ff}} = \sqrt{\frac{\rho_0 r_0^\alpha}{M_\star} r_{\rm cyl}^{3-\alpha}}.
\end{gather}
Here, $\Omega_{\rm K}$ is the Keplerian angular velocity, $G$ is the gravitational constant, and $M_\star$ is the stellar mass. If $\mu_{\rm sg} \ll 1$ the disk is dominated by the gravitational potential of the central star and therefore Keplerian, i.e., it is non-self-gravitating. In contrast, if $\mu_{\rm sg} \gg 1$ the disk is dominated by the gravitational potential of the local mass distribution and so becomes near-Keplerian self-gravitating to fully self-gravitating with increasing $\mu_{\rm sg}$. Figure~\ref{Fig:self_grav} shows the radial dependent ratio $\mu_{\rm sg}$ and it can be seen that in our model the disk of the Butterfly Star is non-self-gravitating at all radii. This also implies that gravitational instabilities in the disk are very unlikely. 
\par
\begin{figure}[h]
 \centering\vspace{-0.25cm}
  \includegraphics[width=0.49\textwidth]{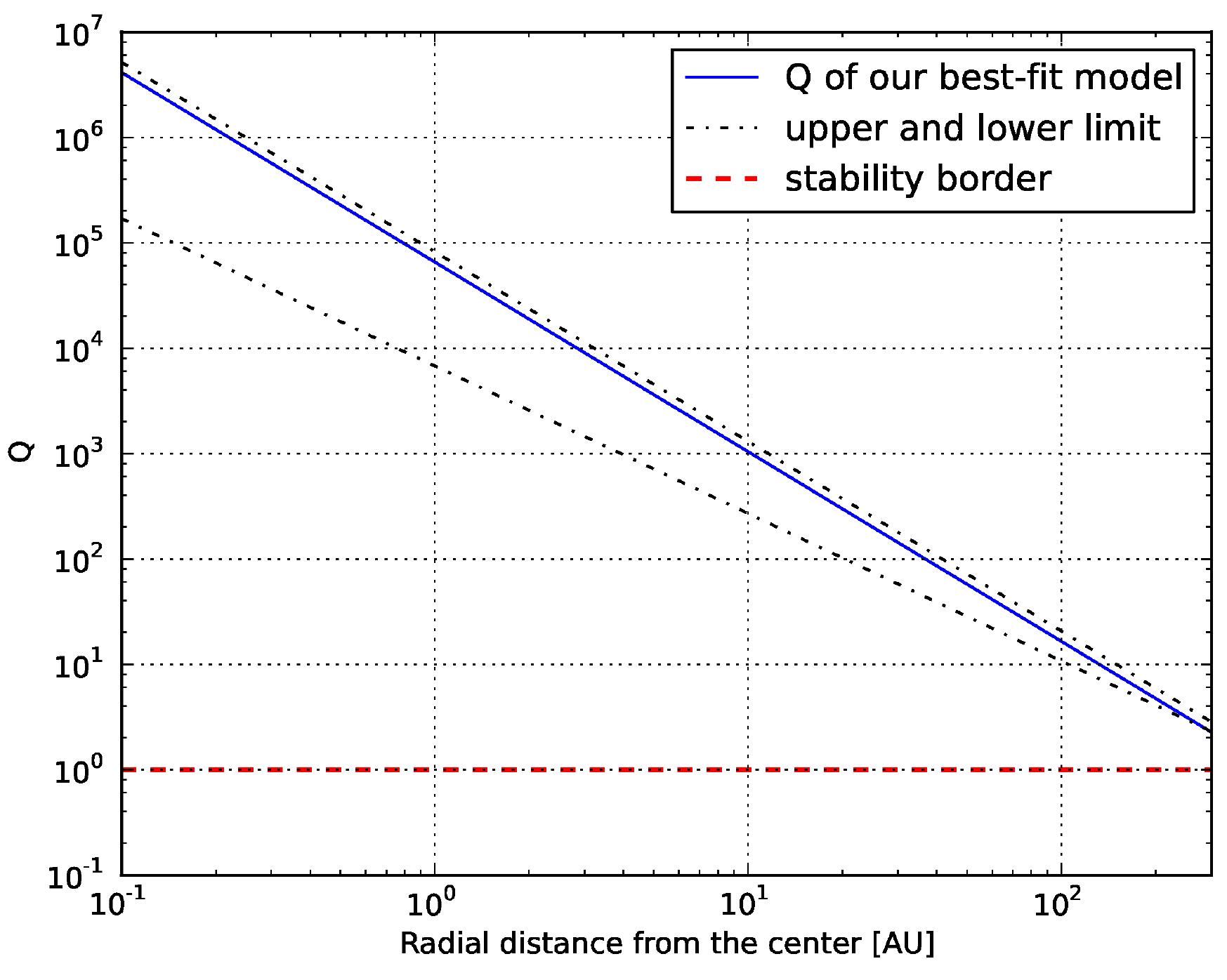}
\caption{Toomre parameter $Q$. The upper and lower limit results from the constraints on the model parameters (see Table~\ref{Tab:overview}).}
\label{Fig:toomre}
\end{figure}

To check this, we made use of the Toomre criterion \citep{Toomre64} that describes when a disk becomes unstable through gravitational collapse. The Toomre $Q$ parameter is defined as
\begin{gather}
Q \equiv \frac{c_{\rm s}\Omega_{\rm K}}{\pi G \Sigma}
\end{gather}
where $c_{\rm s}$ is the local sound speed. In terms of $Q$, a disk is unstable to its own self-gravity if $Q<1$, and stable if $Q > 1$. As Fig.~\ref{Fig:toomre} clearly shows, in our model $Q > 1$ throughout the entire disk, which means that the circumstellar disk of the Butterfly Star is gravitationally stable at all radii as is expected by the determination of $\mu_{\rm sg}$.
\par

  \subsection{Comparison with theoretical models}
  \label{subsection:comparison_theoreticalmodels}

We compared our results qualitatively with theoretical models for the evolution of dust in circumstellar disks by \citet{Garaud04_1}, \citet{Garaud04_2}, \citet{BarriereFouchet05}, and \citet{Garaud07}, for example. The Butterfly Star is characterized as a Class~I YSO and is surrounded by a circumstellar disk that is optically thick even at submm wavelengths and by a substantial envelope. This implies that this object is rather young and potential evolution of dust is in its early phase. The main points of current models on the first phase of planet formation are summarized below. The gas in the protoplanetary disk is partially pressure supported, meaning that both the centrifugal force and the gas pressure counteract the gravity. Because of an outward decreasing gas pressure gradient, the gas orbits at sub-Keplerian velocity around the star. Small dust particles are well-coupled to the gas. As they do not experience the same radial pressure gradient as the gas they orbit with the Keplerian velocity and therefore feel a net inward force causing them to drift inward. As a consequence of processes such as Brownian motion and turbulence in the disk, the dust grains agglomerate as a result of collisions and form aggregates of ever increasing size and mass. In the context of our model we found evidence for the early stage of this grain growth phase in the circumstellar disk of the Butterfly Star with grain radii of up to $\sim\!100\,$\micron. This is significantly larger than what is commonly found in the ISM ($\sim\!0.25\,$\micron). With the increase of the importance of gravity resulting from the continuing grain growth, the particles decouple from the pure gas motion and settle toward the disk midplane. Because they have a faster velocity than the gas does, the larger particles see a headwind that saps their angular momentum and causes them to spiral in toward the star. This radial drift occurs on a time scale that is much shorter than the disk lifetime. From our multi-wavelength modeling of the disk of the Butterfly Star we also found quantitative evidence for the settling of larger dust grains toward the disk midplane while smaller particles are retained in the upper layers of the disk. We found the larger dust grains to be concentrated toward the star which is in agreement with the predicted inward spiraling. Moreover, theoretical models predict that the gas drag efficiency varies according to the grain size, where intermediate-sized particles (100\,\micron\ to 10\,cm) experience the strongest perturbation to their movement. The largest particles that we have found based on our model exhibit a radius of $\sim\!100\,$\micron. Therefore, our findings for radial segregation of the dust is not unexpected. Theoretical work by \citet{Garaud12} suggests the coexistence of two particle populations, one with larger particles and one with smaller particles, at different radial positions in the protoplanetary disk in contrast to a single continuous population from small to large sizes. Our need for two different dust grain-size distributions to properly model the properties of the disk of the Butterfly Star agrees well with this prediction.
\par
We find that our results are consistent with current models for the evolution of dust in circumstellar disks. Our findings of grain growth, vertical settling, and radial segregation in the disk of the Butterfly Star are common predictions of theoretical models.
\par

  \subsection{Observability with ALMA}

\begin{figure*}[!htb]
 \centering
  \includegraphics[width=0.49\textwidth]{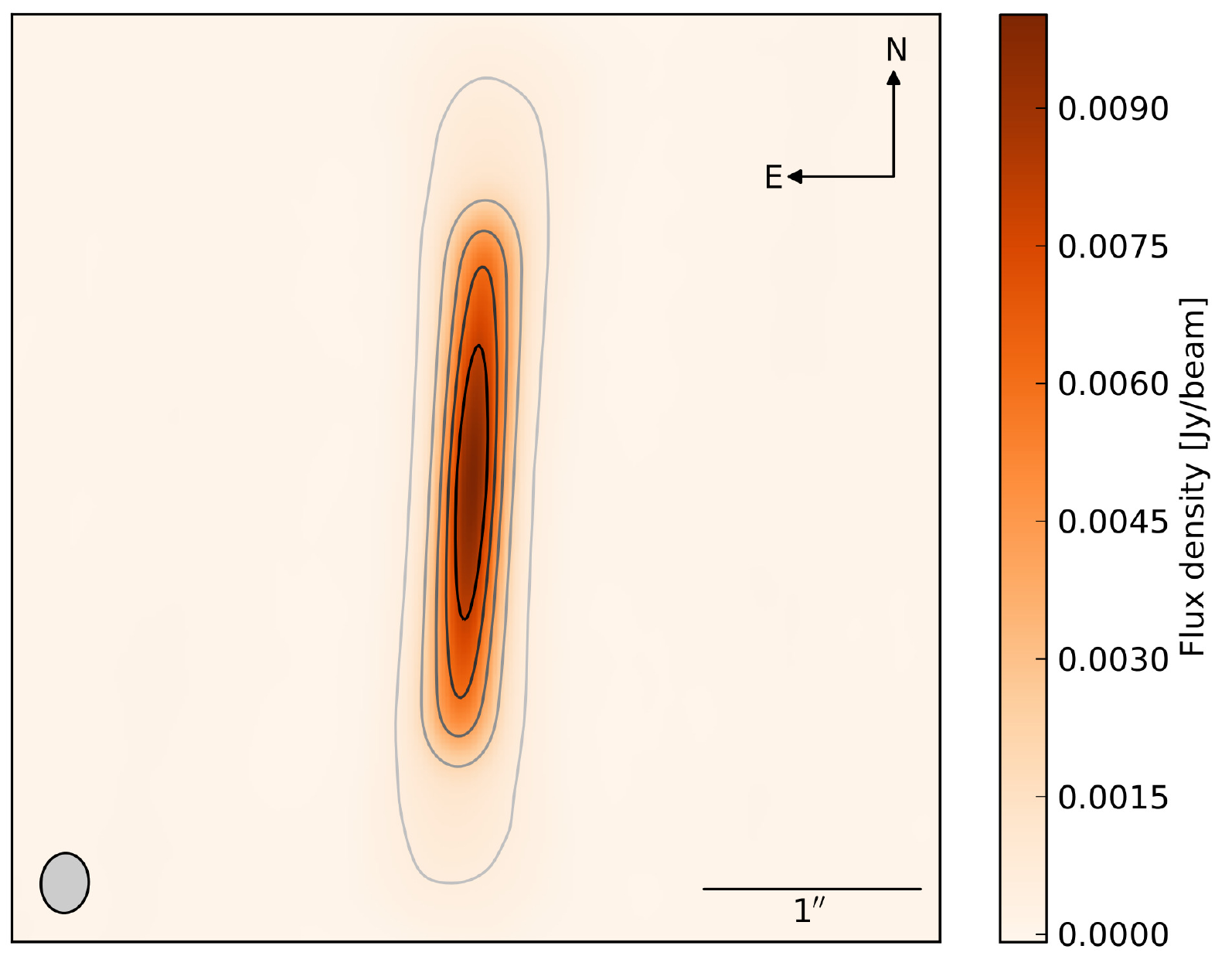}
  \includegraphics[width=0.49\textwidth]{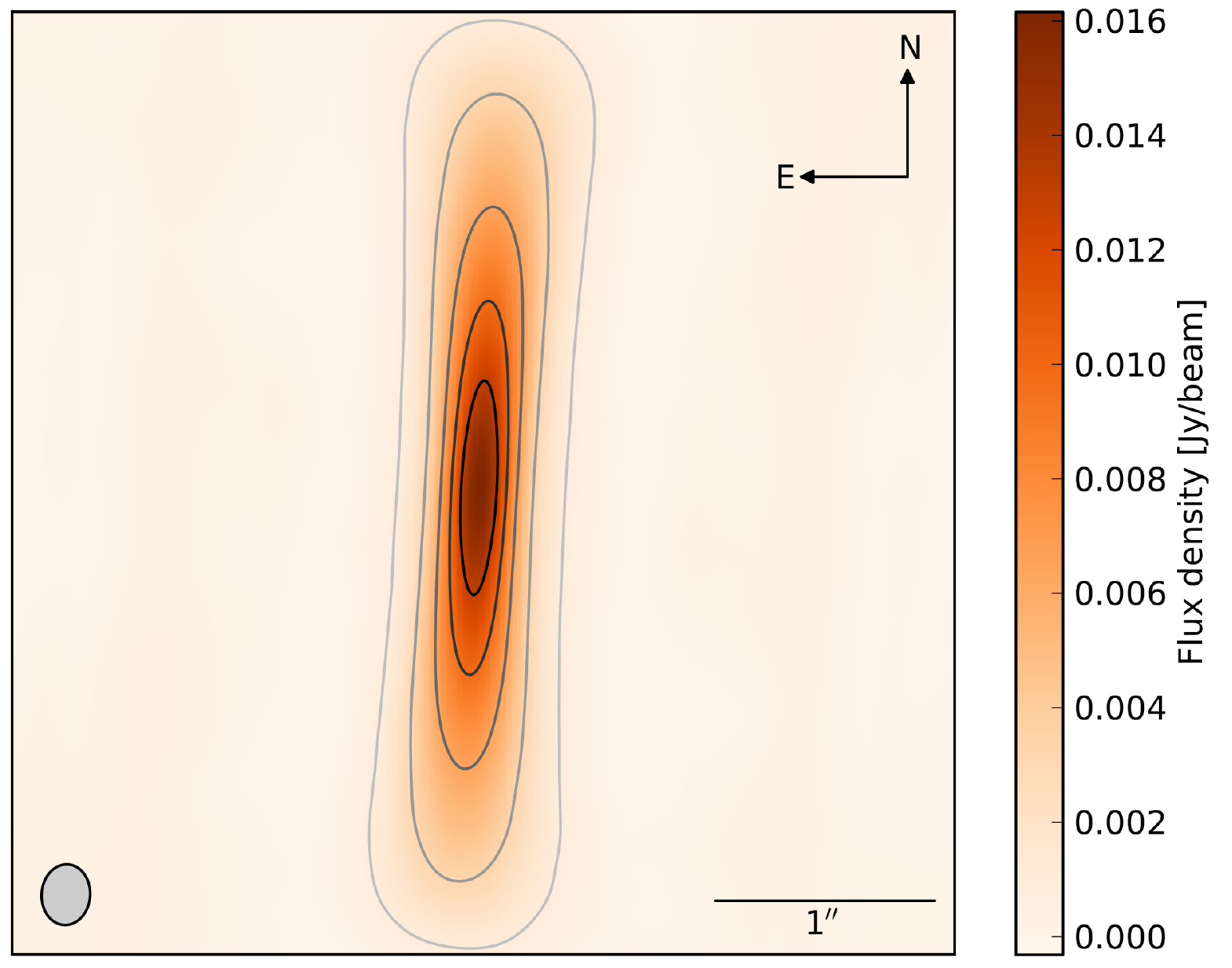}
\caption{Simulated ALMA observation at 1.3\,mm. The contour levels are at 5, 20, 40, 60, and 80\,\% of the maximum value.\newline \emph{Left:} Simulation based on our best-fit model. \emph{Right:} Same as left, but using only one grain-size distribution with $a_{\rm max} = 100\,$\micron.}
\label{Fig:ALMA_mm}
\end{figure*}
\begin{figure*}[!htb]
 \centering
  \includegraphics[width=0.49\textwidth]{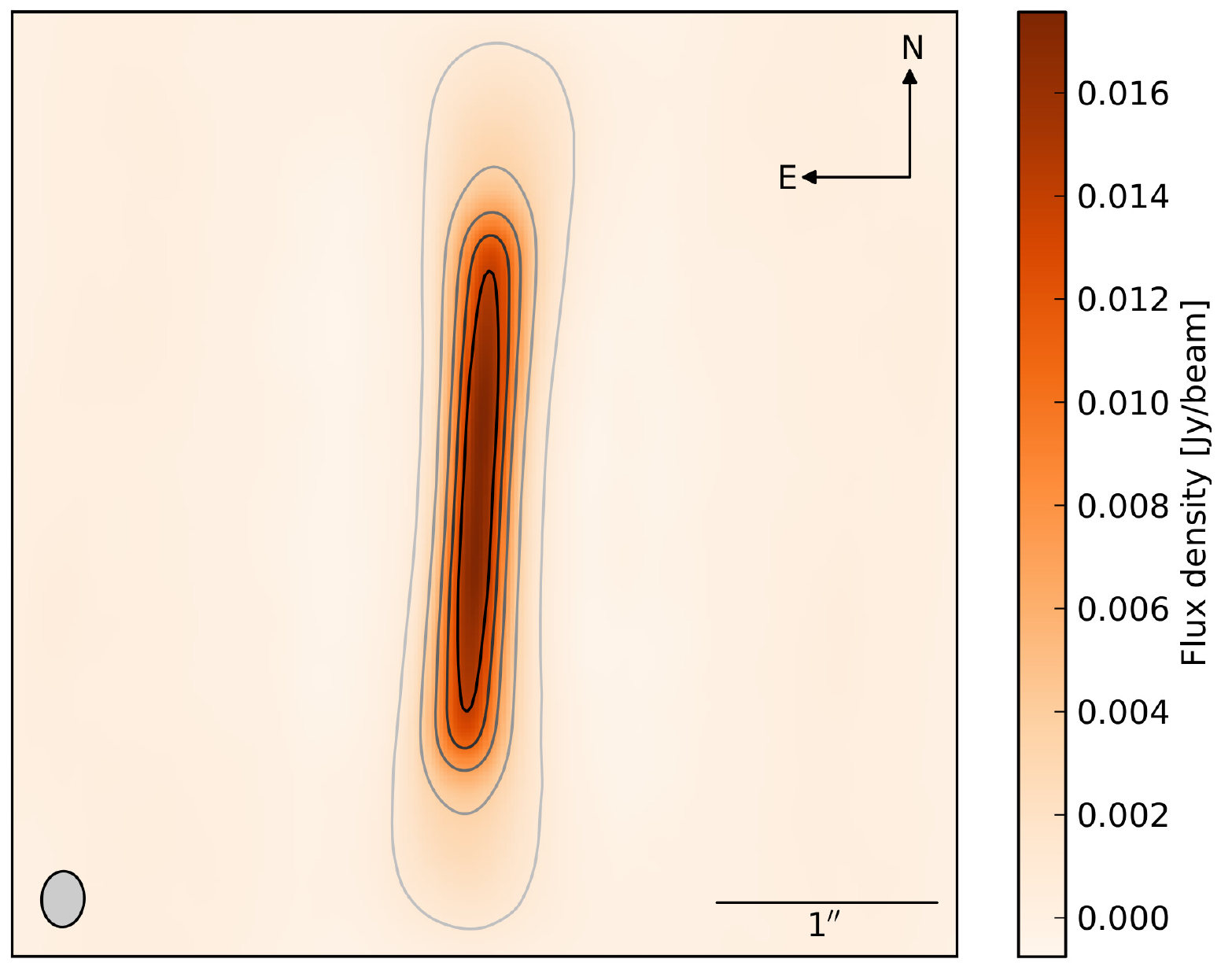}
  \includegraphics[width=0.49\textwidth]{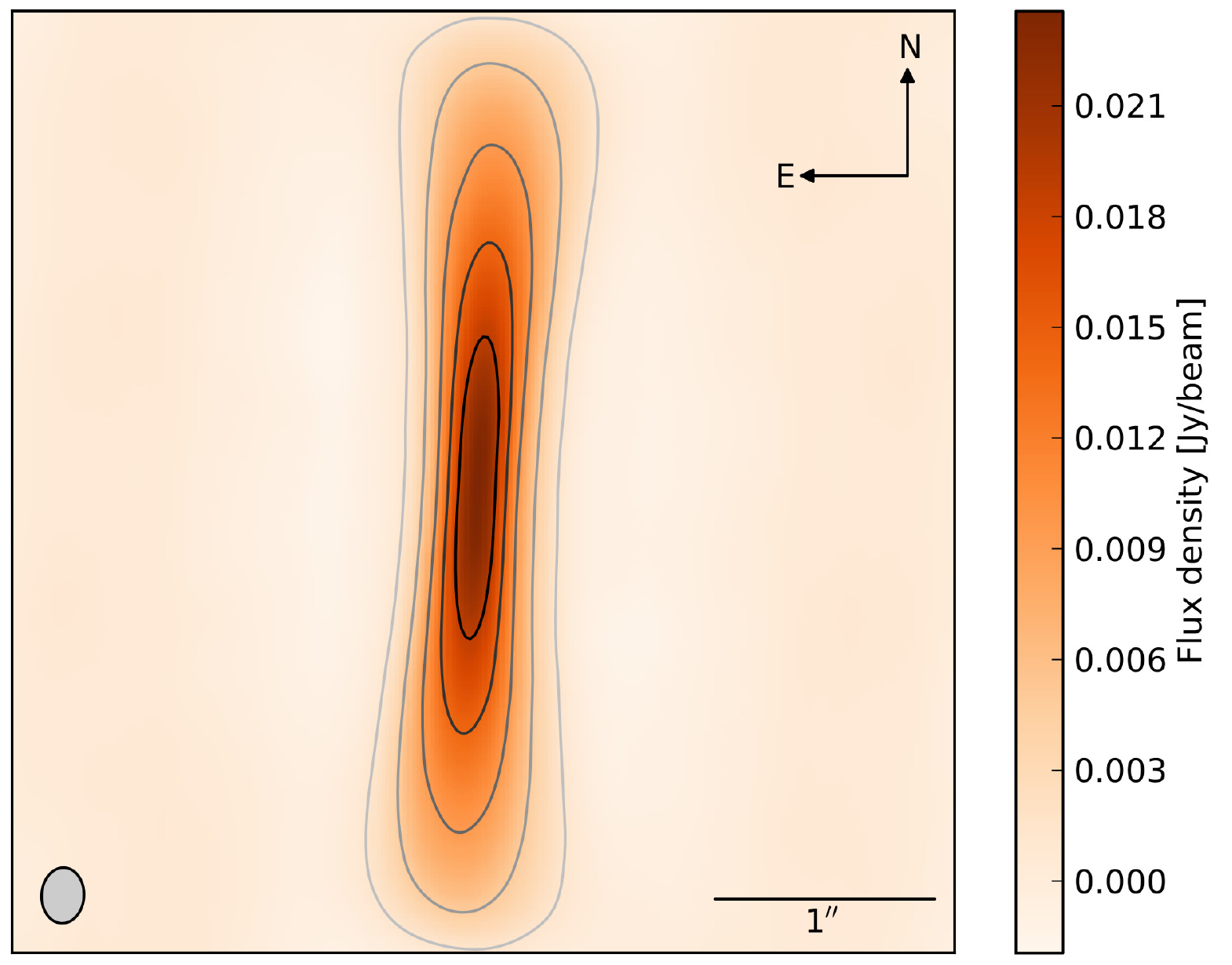}
\caption{Simulated ALMA observation at 894\,\micron. The contour levels are at 5, 20, 40, 60, and 80\,\% of the maximum value.\newline \emph{Left:} Simulation based on our best-fit model. \emph{Right:} Same as left, but using only one grain-size distribution with $a_{\rm max} = 100\,$\micron.}
\label{Fig:ALMA_submm}
\end{figure*}
Figures~\ref{Fig:ALMA_mm} and \ref{Fig:ALMA_submm} show simulated ALMA observations at 1.3\,mm and 894\,\micron, respectively. They are based on the modeled (sub)mm maps of our best-fit and are created using the CASA simulator \citep{CASA08}. Both figures represent the use of the full ALMA array with an observing time of 2 hours, in which Fig.~\ref{Fig:ALMA_mm} reflects configuration 15 and Fig.~\ref{Fig:ALMA_submm} configuration 13. The difference between the simulations shown on the left and the right in both figures is that for the right one only one single dust grain-size distribution with $a_{\rm max} = 100\,$\micron\ is used whereas the simulation on the left exactly represents our best-fit model. The dust re-emission in the region that is encased by the 20\,\% and 40\,\% contour level in Fig.~\ref{Fig:ALMA_mm}, left and Fig.~\ref{Fig:ALMA_submm}, left, respectively, is strongly dominated by the larger dust grains. Because of the much higher angular resolution compared to the SMA and the PdBI, the radial extent of this region is cleary visible. The difference in the radial brightness structure between the presence of only one grain-size distribution or, as we have found in our study, two different grain-size distributions in the circumstellar disk of the Butterfly Star can be clearly seen in both figures. Thus, observations with the full ALMA array will allow us to distinguish between the presence of one or two different dust grain-size distributions and let us check our current model of the disk of the Butterfly Star. Most likely, these observations will allow us to give further constraints on the properties of this prominent circumstellar disk.
\par

\section{Summary and conclusion}
\label{section:summary_conclusion}

We have compiled a high-quality data set for the circumstellar disk of the Butterfly Star, spanning a wide range of wavelengths. We obtained images in the NIR and in the (sub)mm regime, as well as photometric data and a spectrum. We have constructed a detailed model that allows the interpretation of all of these observations with one single set of parameters. A systematic analysis of the parameter space allows us to establish strong constraints on all the parameters of the model. The conclusions we obtain are the following:
\begin{enumerate}
  \item  The disk structure is well constrained. The disk extends from an inner radius of $0.1\,$AU up to 300\,AU. The scale height of the disk is $\sim\!10\,$AU at 100\,AU and varies with a flaring index of $\sim\!1.14$. The surface density exponent is found to be in the range of [0.04, 0.48], which is slightly smaller than the range found in general for other circumstellar disks \citep{Kitamura02, Andrews07, Guilloteau11}. The midplane temperature is determined to be 20\,K at 100\,AU.

  \item The dust mass is found to be $\sim\!9\times 10^{-4}\,{\rm M_{\odot}}$ under the assumption of spherical grains and $\rho_{\rm grain} = 2.5\,{\rm g\,cm^{-3}}$. The dust mass of the region where the large grains are distributed is $\sim\!2.2\times 10^{-4}\,{\rm M_{\odot}}$ and that of the small grains is $\sim\!6.8\times 10^{-4}\,{\rm M_{\odot}}$. With a typical dust-to-gas ratio of 1/100, the total disk mass amounts to $\sim\!9\times 10^{-2}\,{\rm M_{\odot}}$, compared to a mass of the central star of $\sim\!1.7\,{\rm M_{\odot}}$. Based on our constraints on the model parameters, the disk of the Butterfly Star is found to be non-self-gravitating and is, according to the Toomre criterion, gravitationally stable at all radii.

  \item The disk of the Butterfly Star has already entered the first phases of planetary formation. We found quantitative evidence for grain growth, vertical settling, as well as radial segregation in this disk. It is not possible to find a coherent multi-wavelength model of the disk using a single grain-size distribution. To do so, the use of at least two different distributions is necessary. In the outer parts and upper layers of the disk, small ISM-sized dust grains can be found. In the inner disk regions, larger dust grains with a maximum particle radius of up to $\sim\!100\,$\micron\ are located. We constrain this region where the larger grains are distributed to a radial extent of [170, 210]\,AU and a vertical extent of [4, 7]\,AU at 100\,AU. The equivalent scale height of this region amounts to a range of [2.5, 4.3]\,AU and for our best-fit model to $3.7\,$AU which is substantially smaller than the vertical extent and the hydrostatic scale height ($H = 6.9\,$AU), thus providing strong support for vertical settling of the large dust grains in the disk. Our findings of dust evolution in this disk are consistent with current theoretical models.

  \item The millimeter spectral index indicates that the dust grains in the disk midplane have already grown to sizes larger than those found in the ISM, supporting our results.

  \item Although each individual observation provides valuable information on the disk, it is necessary to combine a large set of independent observations (SED and spatially resolved images) from different wavelength regimes in a multi-wavelength study. This approach allowed us to derive qualitatively new conclusions which were not obvious on the basis of individual data sets alone and to strongly reduce model degeneracies.
\end{enumerate}
The large number of observations in different wavelength domains, as well as our coherent multi-wavelength modeling, make the Butterfly Star one of the best studied protoplanetary disks so far. With the upcoming completion of the next-generation interferometer ALMA, observations using this facility will allow us to test our findings of dust evolution, will let us further constrain the radial and vertical structure of this fascinating disk, and will refine our understanding of the evolution of protoplanetary disks.
\par

\begin{acknowledgements}
This work is supported by the DFG through the research group 759 ``The Formation of Planets: The Critical First Growth Phase'' (WO 857/4-2) and project WO 857/12-1. We are grateful to the anonymous referee for providing useful suggestions that greatly improved this paper.
\end{acknowledgements}

\bibliographystyle{aa}
\bibliography{bibfile}

\begin{thebibliography}{90}
\expandafter\ifx\csname natexlab\endcsname\relax\def\natexlab#1{#1}\fi

\bibitem[{{Adams} {et~al.}(1987){Adams}, {Lada}, \& {Shu}}]{AdamsLadaShu87}
{Adams}, F.~C., {Lada}, C.~J., \& {Shu}, F.~H. 1987, \apj, 312, 788

\bibitem[{{Alexander} \& {Armitage}(2007)}]{Alexander07}
{Alexander}, R.~D. \& {Armitage}, P.~J. 2007, \mnras, 375, 500

\bibitem[{{Alexander} \& {Armitage}(2009)}]{Alexander09}
{Alexander}, R.~D. \& {Armitage}, P.~J. 2009, \apj, 704, 989

\bibitem[{{Andrews} \& {Williams}(2005)}]{Andrews05}
{Andrews}, S.~M. \& {Williams}, J.~P. 2005, \apj, 631, 1134

\bibitem[{{Andrews} \& {Williams}(2007)}]{Andrews07}
{Andrews}, S.~M. \& {Williams}, J.~P. 2007, \apj, 659, 705

\bibitem[{{Andrews} {et~al.}(2011){Andrews}, {Wilner}, {Espaillat}, {Hughes},
  {Dullemond}, {McClure}, {Qi}, \& {Brown}}]{Andrews11}
{Andrews}, S.~M., {Wilner}, D.~J., {Espaillat}, C., {et~al.} 2011, \apj, 732,
  42

\bibitem[{{Armitage}(2007)}]{Armitage07}
{Armitage}, P.~J. 2007, Lecture notes on the formation and early evolution of
  planetary systems

\bibitem[{{Armitage}(2010)}]{Armitage10}
{Armitage}, P.~J. 2010, {Astrophysics of Planet Formation}

\bibitem[{{Barri{\`e}re-Fouchet} {et~al.}(2005){Barri{\`e}re-Fouchet},
  {Gonzalez}, {Murray}, {Humble}, \& {Maddison}}]{BarriereFouchet05}
{Barri{\`e}re-Fouchet}, L., {Gonzalez}, J.-F., {Murray}, J.~R., {Humble},
  R.~J., \& {Maddison}, S.~T. 2005, \aap, 443, 185

\bibitem[{{Beckwith} {et~al.}(2000){Beckwith}, {Henning}, \&
  {Nakagawa}}]{Beckwith00}
{Beckwith}, S.~V.~W., {Henning}, T., \& {Nakagawa}, Y. 2000, Protostars and
  Planets IV, 533

\bibitem[{{Bell} {et~al.}(1997){Bell}, {Cassen}, {Klahr}, \&
  {Henning}}]{Bell97}
{Bell}, K.~R., {Cassen}, P.~M., {Klahr}, H.~H., \& {Henning}, T. 1997, \apj,
  486, 372

\bibitem[{{Benz}(2000)}]{Benz00}
{Benz}, W. 2000, \ssr, 92, 279

\bibitem[{{Birnstiel} {et~al.}(2011){Birnstiel}, {Ormel}, \&
  {Dullemond}}]{Birnstiel11}
{Birnstiel}, T., {Ormel}, C.~W., \& {Dullemond}, C.~P. 2011, \aap, 525, A11

\bibitem[{{Bjorkman} \& {Wood}(2001)}]{Bjorkman01}
{Bjorkman}, J.~E. \& {Wood}, K. 2001, \apj, 554, 615

\bibitem[{{Blum} \& {Wurm}(2008)}]{Blum08}
{Blum}, J. \& {Wurm}, G. 2008, \araa, 46, 21

\bibitem[{{Boss}(2002)}]{Boss02}
{Boss}, A.~P. 2002, Earth and Planetary Science Letters, 202, 513

\bibitem[{{Brauer} {et~al.}(2008){Brauer}, {Dullemond}, \&
  {Henning}}]{Brauer08}
{Brauer}, F., {Dullemond}, C.~P., \& {Henning}, T. 2008, \aap, 480, 859

\bibitem[{{Cashwell} \& {Everett}(1959)}]{Cashwell59}
{Cashwell}, E.~D. \& {Everett}, C.~J. 1959, {A practical manual on the Monte
  Carlo method for random walk problems} (Pergamon Press)

\bibitem[{{Chiang} \& {Youdin}(2010)}]{Chiang10}
{Chiang}, E. \& {Youdin}, A.~N. 2010, Annual Review of Earth and Planetary
  Sciences, 38, 493

\bibitem[{{Chiang} \& {Goldreich}(1997)}]{ChiangGoldreich97}
{Chiang}, E.~I. \& {Goldreich}, P. 1997, \apj, 490, 368

\bibitem[{{Chiang} {et~al.}(2001){Chiang}, {Joung}, {Creech-Eakman}, {Qi},
  {Kessler}, {Blake}, \& {van Dishoeck}}]{Chiang01}
{Chiang}, E.~I., {Joung}, M.~K., {Creech-Eakman}, M.~J., {et~al.} 2001, \apj,
  547, 1077

\bibitem[{{Clarke} {et~al.}(2001){Clarke}, {Gendrin}, \&
  {Sotomayor}}]{Clarke01}
{Clarke}, C.~J., {Gendrin}, A., \& {Sotomayor}, M. 2001, \mnras, 328, 485

\bibitem[{{D'Alessio} {et~al.}(2006){D'Alessio}, {Calvet}, {Hartmann},
  {Franco-Hern{\'a}ndez}, \& {Serv{\'{\i}}n}}]{DAlessio06}
{D'Alessio}, P., {Calvet}, N., {Hartmann}, L., {Franco-Hern{\'a}ndez}, R., \&
  {Serv{\'{\i}}n}, H. 2006, \apj, 638, 314

\bibitem[{{Dominik} {et~al.}(2007){Dominik}, {Blum}, {Cuzzi}, \&
  {Wurm}}]{Dominik07}
{Dominik}, C., {Blum}, J., {Cuzzi}, J.~N., \& {Wurm}, G. 2007, Protostars and
  Planets V, 783

\bibitem[{{Draine} \& {Lee}(1984)}]{Draine84}
{Draine}, B.~T. \& {Lee}, H.~M. 1984, \apj, 285, 89

\bibitem[{{Draine} \& {Malhotra}(1993)}]{Draine93}
{Draine}, B.~T. \& {Malhotra}, S. 1993, \apj, 414, 632

\bibitem[{{Dullemond} \& {Dominik}(2004)}]{Dullemond04}
{Dullemond}, C.~P. \& {Dominik}, C. 2004, \aap, 421, 1075

\bibitem[{{Dullemond} \& {Dominik}(2005)}]{Dullemond05}
{Dullemond}, C.~P. \& {Dominik}, C. 2005, \aap, 434, 971

\bibitem[{{Espaillat} {et~al.}(2012){Espaillat}, {Ingleby}, {Hern{\'a}ndez},
  {Furlan}, {D'Alessio}, {Calvet}, {Andrews}, {Muzerolle}, {Qi}, \&
  {Wilner}}]{Espaillat12}
{Espaillat}, C., {Ingleby}, L., {Hern{\'a}ndez}, J., {et~al.} 2012, \apj, 747,
  103

\bibitem[{{Fortier} {et~al.}(2012){Fortier}, {Alibert}, {Carron}, {Benz}, \&
  {Dittkrist}}]{Fortier12}
{Fortier}, A., {Alibert}, Y., {Carron}, F., {Benz}, W., \& {Dittkrist}, K.-M.
  2012, ArXiv e-prints

\bibitem[{{Fromang} \& {Papaloizou}(2006)}]{Fromang06}
{Fromang}, S. \& {Papaloizou}, J. 2006, \aap, 452, 751

\bibitem[{{Garaud}(2007)}]{Garaud07}
{Garaud}, P. 2007, \apj, 671, 2091

\bibitem[{{Garaud} {et~al.}(2004){Garaud}, {Barri{\`e}re-Fouchet}, \&
  {Lin}}]{Garaud04_1}
{Garaud}, P., {Barri{\`e}re-Fouchet}, L., \& {Lin}, D.~N.~C. 2004, \apj, 603,
  292

\bibitem[{{Garaud} \& {Lin}(2004)}]{Garaud04_2}
{Garaud}, P. \& {Lin}, D.~N.~C. 2004, \apj, 608, 1050

\bibitem[{{Garaud} {et~al.}(2012){Garaud}, {Meru}, {Galvagni}, \&
  {Olczak}}]{Garaud12}
{Garaud}, P., {Meru}, F., {Galvagni}, M., \& {Olczak}, C. 2012, ArXiv e-prints

\bibitem[{{Goldreich} \& {Ward}(1973)}]{Goldreich73}
{Goldreich}, P. \& {Ward}, W.~R. 1973, \apj, 183, 1051

\bibitem[{{Gr{\"a}fe} {et~al.}(2011){Gr{\"a}fe}, {Wolf}, {Roccatagliata},
  {Sauter}, \& {Ertel}}]{Graefe11}
{Gr{\"a}fe}, C., {Wolf}, S., {Roccatagliata}, V., {Sauter}, J., \& {Ertel}, S.
  2011, \aap, 533, A89

\bibitem[{{Guilloteau} {et~al.}(2011){Guilloteau}, {Dutrey}, {Pi{\'e}tu}, \&
  {Boehler}}]{Guilloteau11}
{Guilloteau}, S., {Dutrey}, A., {Pi{\'e}tu}, V., \& {Boehler}, Y. 2011, \aap,
  529, A105

\bibitem[{{Haisch} {et~al.}(2001){Haisch}, {Lada}, \& {Lada}}]{Haisch01}
{Haisch}, Jr., K.~E., {Lada}, E.~A., \& {Lada}, C.~J. 2001, \apjl, 553, L153

\bibitem[{{Hartmann} {et~al.}(1998){Hartmann}, {Calvet}, {Gullbring}, \&
  {D'Alessio}}]{Hartmann98}
{Hartmann}, L., {Calvet}, N., {Gullbring}, E., \& {D'Alessio}, P. 1998, \apj,
  495, 385

\bibitem[{{Hillenbrand} \& {White}(2004)}]{Hillenbrand04}
{Hillenbrand}, L.~A. \& {White}, R.~J. 2004, \apj, 604, 741

\bibitem[{{Ho} {et~al.}(2004){Ho}, {Moran}, \& {Lo}}]{SMA04}
{Ho}, P.~T.~P., {Moran}, J.~M., \& {Lo}, K.~Y. 2004, \apjl, 616, L1

\bibitem[{{Hollenbach} {et~al.}(2000){Hollenbach}, {Yorke}, \&
  {Johnstone}}]{Hollenbach00}
{Hollenbach}, D.~J., {Yorke}, H.~W., \& {Johnstone}, D. 2000, Protostars and
  Planets IV, 401

\bibitem[{{Jaeger}(2008)}]{CASA08}
{Jaeger}, S. 2008, in Astronomical Society of the Pacific Conference Series,
  Vol. 394, Astronomical Data Analysis Software and Systems XVII, ed. R.~W.
  {Argyle}, P.~S. {Bunclark}, \& J.~R. {Lewis}, 623

\bibitem[{{Kenyon} {et~al.}(1994){Kenyon}, {Dobrzycka}, \&
  {Hartmann}}]{Kenyon94}
{Kenyon}, S.~J., {Dobrzycka}, D., \& {Hartmann}, L. 1994, \aj, 108, 1872

\bibitem[{{Kitamura} {et~al.}(2002){Kitamura}, {Momose}, {Yokogawa}, {Kawabe},
  {Tamura}, \& {Ida}}]{Kitamura02}
{Kitamura}, Y., {Momose}, M., {Yokogawa}, S., {et~al.} 2002, \apj, 581, 357

\bibitem[{{Kokubo} \& {Ida}(1998)}]{Kokubo98}
{Kokubo}, E. \& {Ida}, S. 1998, \icarus, 131, 171

\bibitem[{{Krist} \& {Hook}(2004)}]{Krist04}
{Krist}, J.~E. \& {Hook}, R.~N. 2004, {The Tiny Tim User's Guide Version 6.3}

\bibitem[{{Krist} {et~al.}(2011){Krist}, {Hook}, \& {Stoehr}}]{Krist11}
{Krist}, J.~E., {Hook}, R.~N., \& {Stoehr}, F. 2011, {20 years of Hubble Space
  Telescope optical modeling using Tiny Tim}

\bibitem[{{Lada}(1987)}]{Lada87}
{Lada}, C.~J. 1987, in IAU Symposium, Vol. 115, Star Forming Regions, ed.
  M.~{Peimbert} \& J.~{Jugaku}, 1--17

\bibitem[{{Lissauer} \& {Stevenson}(2007)}]{Lissauer07}
{Lissauer}, J.~J. \& {Stevenson}, D.~J. 2007, Protostars and Planets V, 591

\bibitem[{{Lucas} \& {Roche}(1997)}]{Lucas97}
{Lucas}, P.~W. \& {Roche}, P.~F. 1997, \mnras, 286, 895

\bibitem[{{Lucas}(1991)}]{PdBI91}
{Lucas}, R. 1991, in Astronomical Society of the Pacific Conference Series,
  Vol.~19, IAU Colloq. 131: Radio Interferometry. Theory, Techniques, and
  Applications, ed. T.~J. {Cornwell} \& R.~A. {Perley}, 449--452

\bibitem[{{Lucy}(1999)}]{Lucy99}
{Lucy}, L.~B. 1999, \aap, 344, 282

\bibitem[{{Lumme} \& {Rahola}(1994)}]{Lumme94}
{Lumme}, K. \& {Rahola}, J. 1994, \apj, 425, 653

\bibitem[{{Madlener} {et~al.}(2012){Madlener}, {Wolf}, {Dutrey}, \&
  {Guilloteau}}]{Madlener12}
{Madlener}, D., {Wolf}, S., {Dutrey}, A., \& {Guilloteau}, S. 2012, \aap, 543,
  A81

\bibitem[{{Mathis} {et~al.}(1977){Mathis}, {Rumpl}, \& {Nordsieck}}]{Mathis77}
{Mathis}, J.~S., {Rumpl}, W., \& {Nordsieck}, K.~H. 1977, \apj, 217, 425

\bibitem[{{Moriarty-Schieven} {et~al.}(1994){Moriarty-Schieven}, {Wannier},
  {Keene}, \& {Tamura}}]{UKT14_94}
{Moriarty-Schieven}, G.~H., {Wannier}, P.~G., {Keene}, J., \& {Tamura}, M.
  1994, \apj, 436, 800

\bibitem[{{Natta} {et~al.}(2007){Natta}, {Testi}, {Calvet}, {Henning},
  {Waters}, \& {Wilner}}]{Natta07}
{Natta}, A., {Testi}, L., {Calvet}, N., {et~al.} 2007, Protostars and Planets
  V, 767

\bibitem[{{Okuzumi}(2009)}]{Okuzumi08}
{Okuzumi}, S. 2009, \apj, 698, 1122

\bibitem[{{Padgett} {et~al.}(2001){Padgett}, {Stapelfeldt}, \&
  {Sargent}}]{Padgett01}
{Padgett}, D., {Stapelfeldt}, K., \& {Sargent}, A. 2001, in Astronomical
  Society of the Pacific Conference Series, Vol. 231, Tetons 4: Galactic
  Structure, Stars and the Interstellar Medium, ed. C.~E. {Woodward}, M.~D.
  {Bicay}, \& J.~M. {Shull}, 586

\bibitem[{{Padgett} {et~al.}(1999){Padgett}, {Brandner}, {Stapelfeldt},
  {Strom}, {Terebey}, \& {Koerner}}]{Padgett99}
{Padgett}, D.~L., {Brandner}, W., {Stapelfeldt}, K.~R., {et~al.} 1999, \aj,
  117, 1490

\bibitem[{{Papaloizou} \& {Terquem}(2006)}]{Papaloizou06}
{Papaloizou}, J.~C.~B. \& {Terquem}, C. 2006, Reports on Progress in Physics,
  69, 119

\bibitem[{{Pinte} {et~al.}(2008){Pinte}, {Padgett}, {M{\'e}nard},
  {Stapelfeldt}, {Schneider}, {Olofsson}, {Pani{\'c}}, {Augereau},
  {Duch{\^e}ne}, {Krist}, {Pontoppidan}, {Perrin}, {Grady}, {Kessler-Silacci},
  {van Dishoeck}, {Lommen}, {Silverstone}, {Hines}, {Wolf}, {Blake}, {Henning},
  \& {Stecklum}}]{Pinte08}
{Pinte}, C., {Padgett}, D.~L., {M{\'e}nard}, F., {et~al.} 2008, \aap, 489, 633

\bibitem[{{Pollack} {et~al.}(1996){Pollack}, {Hubickyj}, {Bodenheimer},
  {Lissauer}, {Podolak}, \& {Greenzweig}}]{Pollack96}
{Pollack}, J.~B., {Hubickyj}, O., {Bodenheimer}, P., {et~al.} 1996, \icarus,
  124, 62

\bibitem[{{Robitaille} {et~al.}(2007){Robitaille}, {Whitney}, {Indebetouw}, \&
  {Wood}}]{Robitaille07}
{Robitaille}, T.~P., {Whitney}, B.~A., {Indebetouw}, R., \& {Wood}, K. 2007,
  \apjs, 169, 328

\bibitem[{{Ruden} \& {Pollack}(1991)}]{Ruden91}
{Ruden}, S.~P. \& {Pollack}, J.~B. 1991, \apj, 375, 740

\bibitem[{{Safronov} \& {Zvjagina}(1969)}]{Safronov69}
{Safronov}, V.~S. \& {Zvjagina}, E.~V. 1969, \icarus, 10, 109

\bibitem[{{Sauter} \& {Wolf}(2011)}]{Sauter11}
{Sauter}, J. \& {Wolf}, S. 2011, \aap, 527, A27

\bibitem[{{Sauter} {et~al.}(2009){Sauter}, {Wolf}, {Launhardt}, {Padgett},
  {Stapelfeldt}, {Pinte}, {Duch{\^e}ne}, {M{\'e}nard}, {McCabe}, {Pontoppidan},
  {Dunham}, {Bourke}, \& {Chen}}]{Sauter09}
{Sauter}, J., {Wolf}, S., {Launhardt}, R., {et~al.} 2009, \aap, 505, 1167

\bibitem[{{Shakura} \& {Sunyaev}(1973)}]{Shakura73}
{Shakura}, N.~I. \& {Sunyaev}, R.~A. 1973, \aap, 24, 337

\bibitem[{{Stapelfeldt} {et~al.}(1998){Stapelfeldt}, {Krist}, {Menard},
  {Bouvier}, {Padgett}, \& {Burrows}}]{Stapelfeldt98}
{Stapelfeldt}, K.~R., {Krist}, J.~E., {Menard}, F., {et~al.} 1998, \apjl, 502,
  L65

\bibitem[{{Stapelfeldt} {et~al.}(2003){Stapelfeldt}, {M{\'e}nard}, {Watson},
  {Krist}, {Dougados}, {Padgett}, \& {Brandner}}]{Stapelfeldt03}
{Stapelfeldt}, K.~R., {M{\'e}nard}, F., {Watson}, A.~M., {et~al.} 2003, \apj,
  589, 410

\bibitem[{{Stark} {et~al.}(2006){Stark}, {Whitney}, {Stassun}, \&
  {Wood}}]{Stark06}
{Stark}, D.~P., {Whitney}, B.~A., {Stassun}, K., \& {Wood}, K. 2006, \apj, 649,
  900

\bibitem[{{Thommes} {et~al.}(2003){Thommes}, {Duncan}, \&
  {Levison}}]{Thommes03}
{Thommes}, E.~W., {Duncan}, M.~J., \& {Levison}, H.~F. 2003, \icarus, 161, 431

\bibitem[{{Thompson} {et~al.}(1998){Thompson}, {Rieke}, {Schneider}, {Hines},
  \& {Corbin}}]{NICMOS98}
{Thompson}, R.~I., {Rieke}, M., {Schneider}, G., {Hines}, D.~C., \& {Corbin},
  M.~R. 1998, \apjl, 492, L95

\bibitem[{{Toomre}(1964)}]{Toomre64}
{Toomre}, A. 1964, \apj, 139, 1217

\bibitem[{{Ubach} {et~al.}(2012){Ubach}, {Maddison}, {Wright}, {Wilner},
  {Lommen}, \& {Koribalski}}]{Ubach12}
{Ubach}, C., {Maddison}, S.~T., {Wright}, C.~M., {et~al.} 2012, \mnras, 425,
  3137

\bibitem[{{Voshchinnikov}(2002)}]{Voshchinnikov02}
{Voshchinnikov}, N.~V. 2002, in Optics of Cosmic Dust, ed. G.~{Videen} \&
  M.~{Kocifaj}, 1

\bibitem[{{Voshchinnikov} {et~al.}(2007){Voshchinnikov}, {Videen}, \&
  {Henning}}]{fluffyparticles07}
{Voshchinnikov}, N.~V., {Videen}, G., \& {Henning}, T. 2007, \ao, 46, 4065

\bibitem[{{Watson} {et~al.}(2007){Watson}, {Stapelfeldt}, {Wood}, \&
  {M{\'e}nard}}]{Watson07}
{Watson}, A.~M., {Stapelfeldt}, K.~R., {Wood}, K., \& {M{\'e}nard}, F. 2007,
  Protostars and Planets V, 523

\bibitem[{{Weidenschilling}(1977)}]{Weidenschilling77}
{Weidenschilling}, S.~J. 1977, \mnras, 180, 57

\bibitem[{{Weingartner} \& {Draine}(2001)}]{Weingartner01}
{Weingartner}, J.~C. \& {Draine}, B.~T. 2001, \apj, 548, 296

\bibitem[{{Wolf}(2003{\natexlab{a}})}]{Wolf03b}
{Wolf}, S. 2003{\natexlab{a}}, \apj, 582, 859

\bibitem[{{Wolf}(2003{\natexlab{b}})}]{Wolf03MC3D}
{Wolf}, S. 2003{\natexlab{b}}, Computer Physics Communications, 150, 99

\bibitem[{{Wolf} {et~al.}(1999){Wolf}, {Henning}, \& {Stecklum}}]{Wolf99}
{Wolf}, S., {Henning}, T., \& {Stecklum}, B. 1999, \aap, 349, 839

\bibitem[{{Wolf} {et~al.}(2003){Wolf}, {Padgett}, \& {Stapelfeldt}}]{Wolf03}
{Wolf}, S., {Padgett}, D.~L., \& {Stapelfeldt}, K.~R. 2003, \apj, 588, 373

\bibitem[{{Wolf} {et~al.}(2008){Wolf}, {Schegerer}, {Beuther}, {Padgett}, \&
  {Stapelfeldt}}]{Wolf08}
{Wolf}, S., {Schegerer}, A., {Beuther}, H., {Padgett}, D.~L., \& {Stapelfeldt},
  K.~R. 2008, \apjl, 674, L101

\bibitem[{{Wright}(1987)}]{Wright87}
{Wright}, E.~L. 1987, \apj, 320, 818

\bibitem[{{Zsom} {et~al.}(2010){Zsom}, {Ormel}, {G{\"u}ttler}, {Blum}, \&
  {Dullemond}}]{Zsom10}
{Zsom}, A., {Ormel}, C.~W., {G{\"u}ttler}, C., {Blum}, J., \& {Dullemond},
  C.~P. 2010, \aap, 513, A57

\end{thebibliography}

\end{document}